\documentclass[10pt, a4paper, onecolumn]{googledeepmind}
\usepackage{graphicx}
\usepackage{amsmath}
\usepackage{booktabs}
\usepackage{longtable}

\usepackage{tabularx}
\renewcommand{\arraystretch}{1.1}

\usepackage{tikz}
\usepackage{hyperref}
\hypersetup{
	colorlinks   = true, 
	urlcolor     = black, 
	linkcolor    = darkgray, 
	citecolor   = gray, 
    breaklinks 	= true
}

\usepackage{xparse}
\usepackage{comment}
\usepackage{algorithm}
\usepackage{algpseudocode}
\usepackage[capitalise]{cleveref}
\usepackage{subcaption}
\usepackage{xcolor}

\newcommand{\Agents}{\mathcal{K}}
\newcommand{\JointActionSpace}{\mathcal{A}}

\newcommand{\ActionSpace}[1]{\mathcal{A}_{#1}}

\newcommand{\JointMdR}{\mu}

\newcommand{\aState}{S}
\newcommand{\jointAction}{A}

\newcommand{\trajectory}[1]{\tau_{#1}}
\newcommand{\agentState}[1]{s_{#1}}
\newcommand{\action}[1]{a_{#1}}

\newcommand{\magnitudeA}{\left| a \right|}
\newcommand{\directionA}{\theta_a}

\newcommand{\deltaMagA}{\delta_{\magnitudeA}}
\newcommand{\deltaDirA}{\delta_{\directionA}}

\newcommand{\actionSubset}[1]{\sigma_{#1,m,d}}

\newcommand{\mdr}[1]{\mu_{#1}}
\newcommand{\validCount}[3]{\mathcal{V}_{#3} \bigl(#1, #2 \bigr)}

\newcommand{\hypervolume}{\mathcal{V}}
\newcommand{\feasibleActionSpace}[1]{\ActionSpace{#1}^\phi}

\newcommand{\nij}{\validCount{\aState}{\jointAction}{j}}
\newcommand{\nijMdR}{\validCount{\aState}{\left[ \jointAction_{i} \leftarrow \mdr{i} \right]}{j}}
\newcommand{\nii}{\validCount{\aState}{\jointAction}{i}}
\newcommand{\niiMdR}{\validCount{\aState}{\left[ \jointAction_{\neg i} \leftarrow \mdr{\neg i} \right]}{i}}

\newcommand{\countNi}[1]{\validCount{\aState}{\jointAction}{#1}}
\newcommand{\countNiMdRj}[2]{\validCount{\aState}{\left[ \jointAction_{#1} \leftarrow \mdr{#1} \right]}{#2}}

\newcommand{\FeAR}{\mathrm{FeAR}}
\newcommand{\clip}[1]{Z \Biggl(#1\Biggr)}

\newcommand{\true}{\mathbf{True}}
\newcommand{\false}{\mathbf{False}}

\newcommand{\mdrB}{\text{Yielding}}
\newcommand{\mdrC}{\text{Unyielding}}
\newcommand{\mdrD}{\text{Yield-Left}}

\newcommand{\Compliant}{\jointAction_{\mdrB}}
\newcommand{\Assertive}{\jointAction_{\mdrC}}
\newcommand{\CompliantLeft}{\jointAction_{\mdrD}}

\newcommand{\fearAM}[4]{\FeAR_{#1,#2}( #3 | #4)}

\newcommand{\fearCalcCaseStudyFigsCircularDiffNr}[3]{%
	\begin{figure*}[t]
		\centering
		\begin{subfigure}[c]{0.33\linewidth}
			\centering
			\begin{subfigure}[c]{\linewidth}
				\centering
			\includegraphics[width=\textwidth, height=0.6\textwidth, keepaspectratio]{Figures/Plots_cFeAR/#1.png}
				\caption{Scenario}
				\vspace{1ex}
				\label{fig:CS#2_scenario}
			\end{subfigure}
			\begin{subfigure}[c]{\linewidth}
				\centering
			\includegraphics[width=1\textwidth, height=0.6\textwidth, keepaspectratio]{Figures/Plots_cFeAR/#1_mdr.png}
				\caption{\textit{Moves de Rigueur}}
				\label{fig:CS#2_scenario_mdr}
			\end{subfigure}
		\end{subfigure}
		\hfill
		\begin{subfigure}[c]{0.3\linewidth}
			\begin{subfigure}[t]{0.52\linewidth}
				\raggedleft
				\captionsetup{justification=raggedleft}
				\caption*{\tiny{$\quad\validCount{\aState}{\jointAction}{2}$}}
				\includegraphics[trim=0 22 0 0, clip, width=0.95\textwidth]{Figures/Plots_cFeAR/#1_move_actionSpace_1-2.png}
				\label{fig:CS#2_FeAS_a1}
			\end{subfigure}
			\hfill
			\begin{subfigure}[t]{0.46\linewidth}
				\raggedleft
				\caption*{\tiny{$\quad\validCount{\aState}{\jointAction}{1}$}}
				\includegraphics[trim=20 22 0 0, clip, width=0.95\textwidth]{Figures/Plots_cFeAR/#1_move_actionSpace_2-1.png}
				\label{fig:CS#2_FeAS_a2}
			\end{subfigure}
			\hfill
			\begin{subfigure}[t]{0.52\linewidth}
                \captionsetup{justification=centering}
				\caption*{\tiny{$\validCount{\aState}{\left[ \jointAction_{1}\leftarrow \mdr{1} \right]}{2}$}}
				\includegraphics[trim=0 0 0 0, clip, width=0.95\textwidth]{Figures/Plots_cFeAR/#1_mdr_actionSpace_1-2.png}
				\label{fig:CS#2_FeAS_mdr1}
			\end{subfigure}
			\hfill
			\begin{subfigure}[t]{0.46\linewidth}
				\captionsetup{justification=centering}
				\caption*{\tiny{$\ \validCount{\aState}{\left[ \jointAction_{2} \leftarrow \mdr{2} \right]}{1}$}}
				\includegraphics[trim=20 0 0 0, clip, width=0.95\textwidth]{Figures/Plots_cFeAR/#1_mdr_actionSpace_2-1.png}
				\label{fig:CS#2_FeAS_mdr2}
			\end{subfigure}
			\captionsetup{justification=centering}
			\vspace{-2pt}
			\caption{Feasible action space.}
			\label{fig:CS#2_FeAS}
		\end{subfigure}
		\hfill
		\begin{subfigure}[c]{0.3\linewidth}
			\centering
			\includegraphics[width=\textwidth]{Figures/Plots_cFeAR/#1_fear.png}
			\captionsetup{justification=centering}
			\caption{FeAR Values}
			\label{fig:CS#2_FeAR}
		\end{subfigure}
		\begin{subfigure}[c]{\linewidth}
			\centering
			\footnotesize
			\begin{minipage}[c]{0.8\linewidth}
				\centering
				\begin{alignat}{7}
					\FeAR_{1,2} &= \frac{\countNiMdRj{1}{2}-\countNi{2}}{\countNiMdRj{1}{2}} &=& \frac{\raisebox{-0.5ex}{\includegraphics[width=0.35cm]{Figures/Plots_cFeAR/#1_mdr_actionSpace_1-2_stripped.png}}-\raisebox{-0.5ex}{\includegraphics[width=0.35cm]{Figures/Plots_cFeAR/#1_move_actionSpace_1-2_stripped.png}}}{\raisebox{0.5ex}{\includegraphics[width=0.35cm]{Figures/Plots_cFeAR/#1_mdr_actionSpace_1-2_stripped.png}}} \notag\quad
					&\FeAR_{2,2} &= \frac{\countNi{2}}{\countNiMdRj{1}{2}} &=& \frac{\raisebox{-0.5ex}{\includegraphics[width=0.35cm]{Figures/Plots_cFeAR/#1_move_actionSpace_1-2_stripped.png}}}{\raisebox{0.5ex}{\includegraphics[width=0.35cm]{Figures/Plots_cFeAR/#1_mdr_actionSpace_1-2_stripped.png}}} \notag \\
					\FeAR_{1,1}  &= \frac{\countNi{1}}{\countNiMdRj{2}{1}} &=&
					\frac{\raisebox{-0.5ex}{\includegraphics[width=0.35cm]{Figures/Plots_cFeAR/#1_move_actionSpace_2-1_stripped.png}}}{\raisebox{0.5ex}{\includegraphics[width=0.35cm]{Figures/Plots_cFeAR/#1_mdr_actionSpace_2-1_stripped.png}}} \notag\quad
					&\FeAR_{2,1} &=
					\frac{\countNiMdRj{2}{1}-\countNi{1}}{\countNiMdRj{2}{1}} &=&
					\frac{\raisebox{-0.5ex}{\includegraphics[width=0.35cm]{Figures/Plots_cFeAR/#1_mdr_actionSpace_2-1_stripped.png}} - \raisebox{-0.5ex}{\includegraphics[width=0.35cm]{Figures/Plots_cFeAR/#1_move_actionSpace_2-1_stripped.png}}}{\raisebox{0.5ex}{\includegraphics[width=0.35cm]{Figures/Plots_cFeAR/#1_mdr_actionSpace_2-1_stripped.png}}} \notag
				\end{alignat}
			\end{minipage}
			\normalsize
			\caption{Calculating $\FeAR_{i,j}$.}
			\label{fig:CS#2_FeAR_calc}
		\end{subfigure}
		\caption{#3}
		\label{fig:CaseStudy#2}
	\end{figure*}
}

\newcounter{casestudy}

\crefformat{casestudy}{Fig.~\thefigure#2#1#3}
\Crefformat{casestudy}{Fig.~\thefigure#2#1#3}
\crefrangeformat{casestudy}{Figs.~\thefigure#3#1#4 to~\thefigure#5#2#6}
\Crefrangeformat{casestudy}{Figs.~\thefigure#3#1#4 to~\thefigure#5#2#6}
\crefmultiformat{casestudy}{Figs.~\thefigure#2#1#3}{ and~\thefigure#2#1#3}{, \thefigure#2#1#3}{ and~\thefigure#2#1#3}
\Crefmultiformat{casestudy}{Figs.~\thefigure#2#1#3}{ and~\thefigure#2#1#3}{, \thefigure#2#1#3}{ and~\thefigure#2#1#3}

\newcommand{\CaseStudySubFigs}[3]{%
	\hfill
	\begin{minipage}[c]{0.98\linewidth}
		\refstepcounter{casestudy}%
		\label{fig:CaseStudy#2}%
		\makebox[0.05\linewidth][c]{(\alph{casestudy})}%
		\hfill
		\begin{minipage}[c]{0.6\linewidth}
			\centering
			\includegraphics[width=\textwidth, height=0.45\textwidth, 
			keepaspectratio]{Figures/Plots_cFeAR/#1.png}
		\end{minipage}%
		\begin{minipage}[c]{0.3\linewidth}
			\centering
			\includegraphics[width=\linewidth]{Figures/Plots_cFeAR/#1_fear.png}
		\end{minipage}
	\end{minipage}
}

\newcommand{\MdRCaseStudySubFigsBarsPicked}[3]{%
	\begin{subfigure}[c]{\linewidth}
		\centering
		\vspace{0.5ex}
		\begin{subfigure}[c]{0.3\linewidth}
			\centering
			\begin{tikzpicture}
				\draw[draw=white] (0, 0) rectangle (\textwidth, 0.8\textwidth);
				\node[anchor=center] at (\textwidth/2, 0.8\textwidth/2) {
					\includegraphics[width=\textwidth, height=\textwidth, keepaspectratio]{Figures/Plots_cFeAR/#1_scenario_0.png}
				};
			\end{tikzpicture}
			\caption{ $\jointAction = \jointAction_{\csname mdr#2\endcsname}$}
			\label{fig:CaseStudyMdR#2}
		\end{subfigure}
		\begin{minipage}{0.05\linewidth}
			\centering
			\rotatebox{90}{\footnotesize FeAR values}
		\end{minipage}
		\hfill
		\begin{subfigure}[c]{0.2\linewidth}
			\centering
			\includegraphics[width=\linewidth]{Figures/Plots_cFeAR/#1_scenario_1_fear_bars_panel.png}
		\end{subfigure}
		\hfill
		\begin{subfigure}[c]{0.2\linewidth}
			\centering
			\includegraphics[width=\linewidth]{Figures/Plots_cFeAR/#1_scenario_2_fear_bars_panel.png}
		\end{subfigure}
		\hfill
		\begin{subfigure}[c]{0.2\linewidth}
			\centering
			\includegraphics[width=\linewidth]{Figures/Plots_cFeAR/#1_scenario_3_fear_bars_panel.png}
		\end{subfigure}
		\\ \hrulefill
	\end{subfigure}
}

\newcommand{\backwardResponsibilityFigs}[3]{%
\begin{figure*}[ht]
    \centering
    \begin{subfigure}[c]{0.99\linewidth}
     \centering
    \begin{subfigure}[c]{0.45\linewidth}
    \centering
    \includegraphics[width=\textwidth]{Figures/Plots_cFeAR/#2_scenario_0.png}
    \caption{Scenario}
    \label{fig:BackwardFear_#1_scenario}
    \end{subfigure}
\hfill
    \begin{subfigure}[c]{0.45\linewidth}
    \centering
    \includegraphics[width=\textwidth]{Figures/Plots_cFeAR/#2_scenario_0_mdr.png}
    \caption{The \textit{Moves de Rigueur}}
    \label{fig:BackwardFear_#1_mdr}
    \end{subfigure}
\hfill
    \begin{subfigure}[c]{0.45\linewidth}
    \centering
    \includegraphics[width=\textwidth]{Figures/Plots_cFeAR/#2_scenario_0_fear.png}
    \caption{FeAR values}
    \label{fig:BackwardFear_#1_fear}
    \end{subfigure}
\hfill
    \begin{subfigure}[c]{0.5\linewidth}
    \centering
    \includegraphics[width=\textwidth]{Figures/Plots_cFeAR/#2_fearGraph_0.png}
    \caption{Directed graph showing the FeAR values}
    \label{fig:BackwardFear_#1_fear_graph}
    \end{subfigure}
    \end{subfigure}
    \caption{#3}
    \label{fig:BackwardFear_#1}
\end{figure*}
}

\newcommand{\forwardResponsibilityFigs}[4]{%
\begin{figure*}[htb]
         \centering
    \begin{subfigure}[b]{\linewidth}
    \centering
    \includegraphics[width=\textwidth]{Figures/Plots_cFeAR/#2_fear_grids_actor_#3_each_affected.png}
    \captionsetup{justification=centering}
    \caption{FeAR values of Agent 3 for each action per affected agent}
    \label{fig:ForwardFear_#1_gridsearch_affected}
    \end{subfigure}
\hfill
    \begin{subfigure}[c]{0.6\linewidth}
    \begin{subfigure}[b]{\linewidth}
    \centering
    \includegraphics[width=\textwidth]{Figures/Plots_cFeAR/#2_fear_grids_actor_#3_aggregates.png}
    \captionsetup{justification=centering}
    \caption{Aggregate FeAR values of Agent 3 across affected agents.}
    \label{fig:ForwardFear_#1_gridsearch_aggregates}
    \end{subfigure}
\hfill
    \begin{subfigure}[b]{\linewidth}
    \centering
    \includegraphics[width=\textwidth]{Figures/Plots_cFeAR/#2_fear_grids_actor_#3_aggregates_counts.png}
    \captionsetup{justification=centering}
    \caption{Number of agents that agent 3 is being courteous or assertive to.}
    \label{fig:ForwardFear_#1_gridsearch_aggregates_counts}
    \end{subfigure}
    \end{subfigure}
\hfill
    \begin{subfigure}[c]{0.39\linewidth}
    \centering
    \includegraphics[trim=0 0 0 20, clip, width=0.9\textwidth]{Figures/Plots_cFeAR/#2_actor_#3_mean_fearless_actions.png}
    \captionsetup{justification=centering}
    \caption{The optimal action of Agent 3 that minimizes mean FeAR}
    \label{fig:ForwardFear_#1_Optimal_action}
    \end{subfigure}
    \caption{#4}
    \label{fig:ForwardLookingResponsibility_#1}
\end{figure*}
}

\begin{document}

\title{Feasible Action Space Reduction for Quantifying Causal Responsibility in Continuous Spatial Interactions}

\author[1]{Ashwin George}
\author[1]{Luciano Cavalcante Siebert}
\author[1]{David Abbink}
\author[1]{Arkady Zgonnikov}

\affil[1]{Deflt University of Technology}

\begin{abstract}
Understanding the causal influence of one agent on another agent is crucial for safely deploying artificially intelligent systems such as automated vehicles and mobile robots into human-inhabited environments.
Existing models of causal responsibility deal with simplified abstractions of scenarios with discrete actions, thus, limiting real-world use when understanding responsibility in spatial interactions. 
Based on the assumption that spatially interacting agents are embedded in a scene and must follow an action at each instant, Feasible Action-Space Reduction (FeAR) was proposed as a metric for causal responsibility in a grid-world setting with discrete actions.
Since real-world interactions involve continuous action spaces, this paper proposes a formulation of the FeAR metric for measuring causal responsibility in space-continuous interactions.
We illustrate the utility of the metric in prototypical space-sharing conflicts, and showcase its applications for analysing backward-looking responsibility and in estimating forward-looking responsibility to guide agent decision making.
Our results highlight the potential of the FeAR metric for designing and engineering artificial agents, as well as for assessing the responsibility of agents around humans.
\end{abstract}

\maketitle


\section{Introduction}
\label{sec:Introduction}

Automated agents interacting with humans in safety-critical scenarios must behave responsibly and in accordance with ethical principles~\cite{dignumResponsibleArtificialIntelligence2019,dastaniResponsibilityAISystems2023,europeancommission.directorategeneralforcommunicationsnetworkscontentandtechnology.EthicsGuidelinesTrustworthy2019}.  
Measuring how responsibility is distributed 
in such human-agent interactions is crucial for the users, designers, policy-makers, and regulators of automated systems to ensure that these systems are beneficial for the society~\cite{beckersDriversPartiallyAutomated2022, papadimitriouCommonEthicalSafe2022, santonidesioEuropeanCommissionReport2021, calvertDesigningAutomatedVehicle2023}.
In particular, if artificial agents could quantify responsibility, they would be able to reason better about their behaviours in terms of responsibilities towards other agents \cite{dignum2020agents,yazdanpanahReasoningResponsibilityAutonomous2022}.
Backward-looking responsibility, which concerns the responsibility of agents for past actions, is useful for reinforcing desirable behaviours of automated agents. Forward-looking responsibility, which concerns responsibility of agents for future actions, is useful for automated agents while planning actions. 
Among multiple established notions of responsibility~\cite{vincentStructuredTaxonomyResponsibility2011}, \textit{causal responsibility} is especially relevant for human-agent interaction. 
Specifically, causal responsibility can be a prerequisite for other forms of responsibility like liability, blame, praise, and moral responsibility \cite{hartCausationLaw2002,vincentStructuredTaxonomyResponsibility2011, vandepoelEthicsTechnologyEngineering2011}, so quantifying it can aid both humans and artificial agents to reason about responsibility. In this work, we aim to define a metric for causal responsibility that is applicable to spatial interactions. 

Responsibility-aware navigation in mixed traffic (where humans and automated vehicles share the same space) has garnered much interest in recent years \cite{shalev-shwartzFormalModelSafe2018,cosnerLearningResponsibilityAllocations2023, geisslingerEthicalTrajectoryPlanning2023}.
Works on responsibility sensitive navigation have incorporated a term for responsibility in the cost function of the automation which is either derived from traffic rules, physical constraints and risk to other road users \cite{geisslingerEthicalTrajectoryPlanning2023}, or is learned from data  \cite{cosnerLearningResponsibilityAllocations2023,remyLearningResponsibilityAllocations2024}. These pragmatic approaches to model responsibility are rooted in existing traffic rules and how humans currently behave in spatial interactions, and, the introduction of automated driving systems with different sensing, cognitive and actuation capabilities necessitate the creation of a new ethics for transportation \cite{santonidesioEuropeanCommissionReport2021}.
In this direction, the responsibility-sensitive safety (RSS) approach~\cite{shalev-shwartzFormalModelSafe2018, shalev-shwartzVisionZeroProvable2019} formalises rules based on ``duty of care'' and prescribes ``proper responses'' for all vehicles, which, if followed would ensure safety. As per RSS, agents that deviate from their ``proper response'' are held responsible for collisions. These models of responsibility do not explicitly consider causality which is an important condition for ascribing responsibility, especially in legal settings~\cite{hartCausationLaw2002}. As such, there is a need for fundamentally rooted models of responsibility that can be applied to spatial navigation.

In the domain of formal reasoning, several computational frameworks relating to 1) the enforcement of an outcome, 2) being pivotal for an outcome or, 3) affecting the probability of an outcome have been used to operationalise causal responsibility.
Logical frameworks related to the capacity of agents and groups of agents to cause or preclude an outcome have been used to evaluate whether agents are causally responsible for outcomes \cite{loriniLogicalAnalysisResponsibility2014,duijfLogicalStudyMoral2023, gladyshevGroupResponsibilityExceeding2023} and to determine to what degree they are responsible~\cite{yazdanpanahApplyingStrategicReasoning2021,yazdanpanahDistantGroupResponsibility2016}.
Another approach to model causal responsibility uses counterfactual reasoning to discern whether an agent's action is pivotal for an outcome --- that is, whether the outcome would have happened if the agent had not performed the action~\cite{englTheoryCausalResponsibility2018, halpernCausesExplanationsStructuralModel2005}.
Some of these approaches also use probabilistic reasoning about how distant an agent's action is from being pivotal for an outcome to quantify responsibility of agents and groups of agents~\cite{englTheoryCausalResponsibility2018,bartlingShiftingBlameDelegation2012, chocklerResponsibilityBlameStructuralModel2004,halpernCauseResponsibilityBlame2015,triantafyllouActualCausalityResponsibility2022,alechina2017causality}.
Furthermore, specifically in the context of interactions between humans and artificial agents, information-theoretic metrics have been used to quantify causal responsibility of humans 
\cite{douerResponsibilityQuantificationModel2020, douerJudgingOneOwn2022}.
The aforementioned formal methods can deal with general descriptions of events, 
but this expressivity comes at a higher computational cost which limits their application to relatively simple cases with few actions and states. Spatial interactions, on the other hand, involve continuous action and state spaces, which necessitates the formulation of a new metric for causal responsibility.

Thus, there is a gap between the more fundamental but computationally expensive models of causal responsibility and the more pragmatic approaches to model responsibility in spatial interactions that lack fundamental rigour. A recent approach to bridge this gap proposed \emph{\textbf{Fe}asible \textbf{A}ction-Space \textbf{R}eduction (\textbf{FeAR})} as a metric for causal responsibility in spatial interactions~\cite{georgeFeasibleActionSpaceReduction2023a}. However, that work is limited to toy scenarios with discrete and finite action spaces, restricting its applicability to real-life human-agent interactions\footnote{For a detailed discussion on existing models of responsibility, please see \cref{sec:Discussion}}.

\noindent\textbf{Contributions:}
Here we (i)~\textit{develop the Feasible Action-Space Reduction (FeAR) metric for causal responsibility in spatial interactions with continuous action spaces}, (ii)~\textit{evaluate the properties of the FeAR metric in case studies representing prototypical space-sharing conflicts}, and (iii)~\textit{illustrate the application of FeAR for analysing backward-looking responsibility in past interactions and estimating forward-looking responsibility for future actions of agents}.

\section{Feasible Action-Space Reduction (FeAR)}

\begin{figure*}[ht]
    \centering
    \begin{subfigure}[c]{0.25\linewidth}
    \centering
    \includegraphics[height=0.95\textwidth]{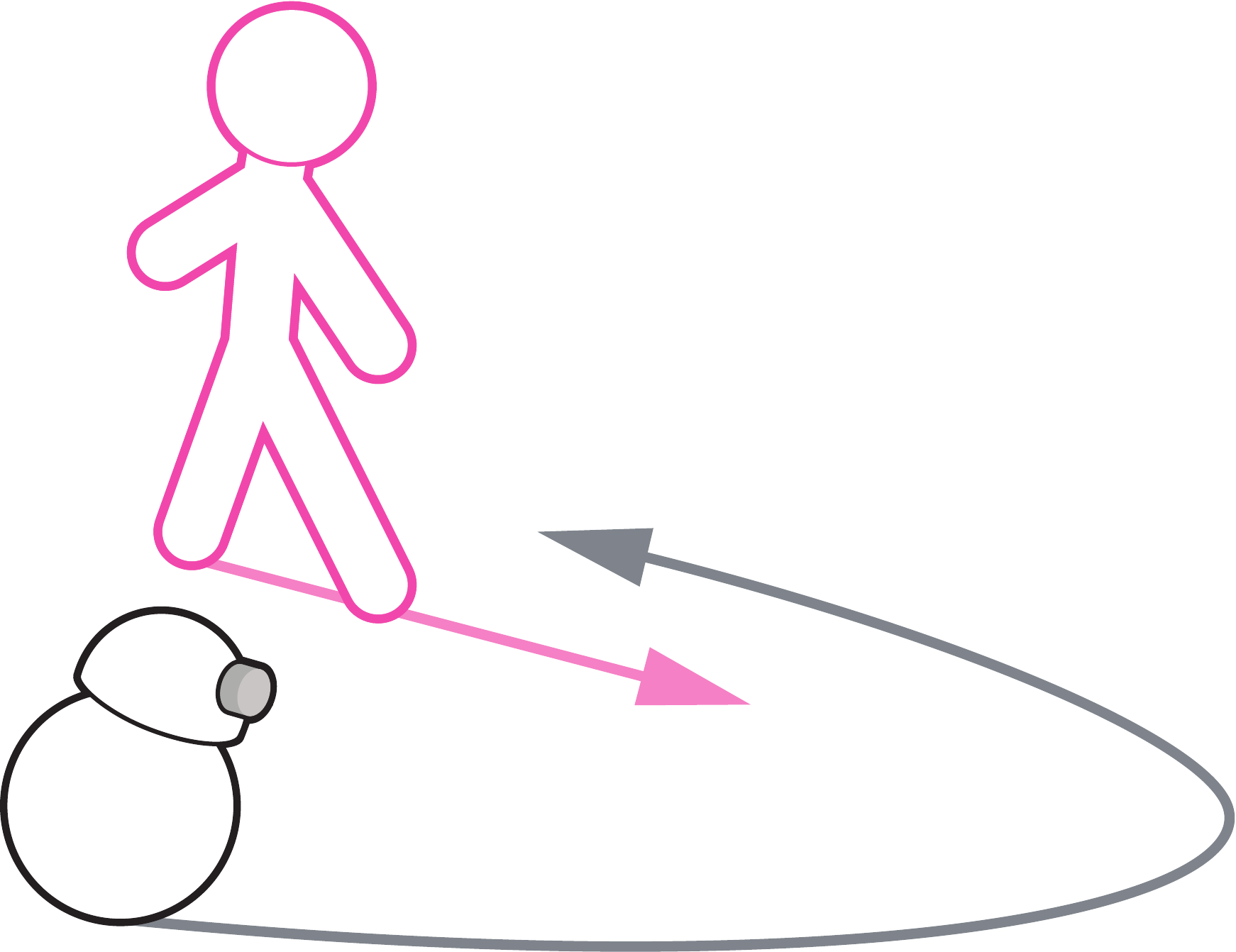}
    \caption{The interaction.}
    \label{fig:cFeAR_Scenario_illustration}
    \end{subfigure}
    \hfill
    \begin{subfigure}[c]{0.34\linewidth}
    \centering
    \includegraphics[height=0.7\linewidth]{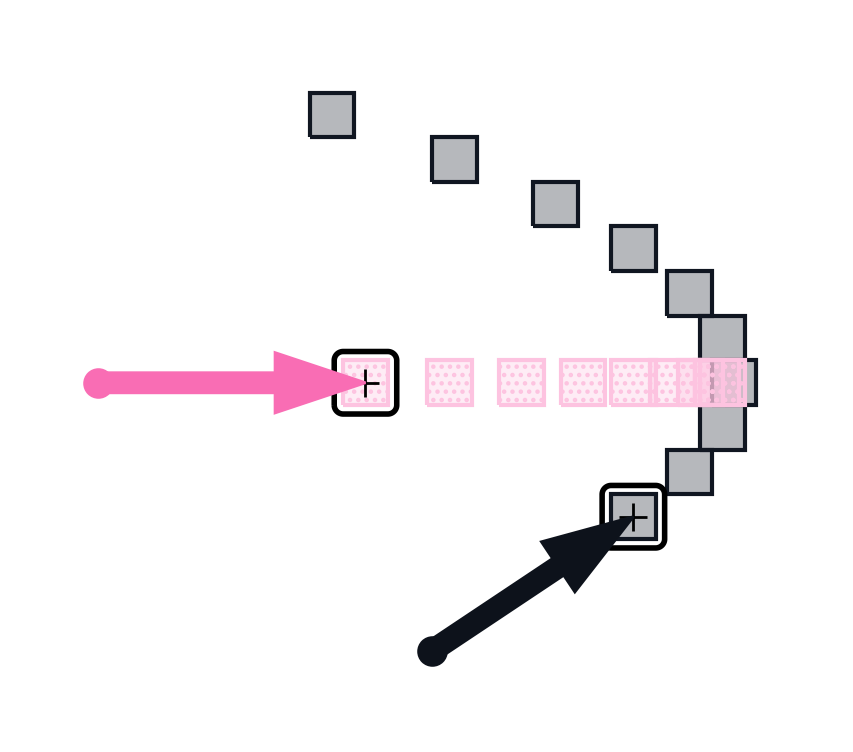}
    \caption{Bounding boxes at each time step.}
    \label{fig:cFeAR_Scenario_boxes}
    \end{subfigure}
    \hfill
    \begin{subfigure}[c]{0.34\linewidth}
    \centering
    \includegraphics[height=0.7\textwidth]{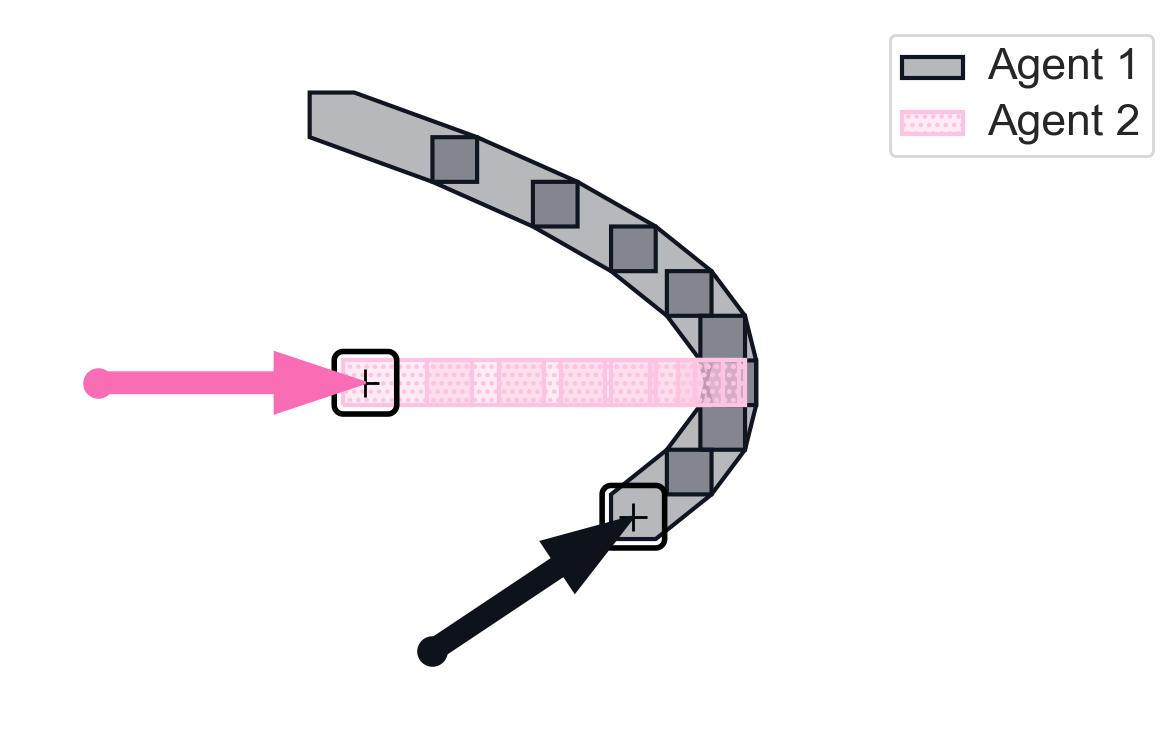}
    \caption{Convex hulls for each time interval.}
    \label{fig:cFeAR_Scenario_hulls}
    \end{subfigure}
\caption{\textbf{Representing a spatial interaction } --- 
(a) a space-sharing interaction where agent 1 (the robot, grey) cuts off agent 2 (the human, pink) while doing a U-Turn;
(b) constant acceleration trajectories $\widehat{\trajectory{i}}$ of the agents at each time step within a time window $[0, T]$. The starting locations of agents $(x_{i}(0), y_{i}(0))$ are indicated by the black outlines, and the initial velocities of agents $(v_{x,i}(0), v_{y,i}(0))$ are indicated by the lengths of the arrows;
(c) convex hulls of the bounding boxes at the start ($t$) and end ($t+ \delta t$) of  time intervals $\Delta T_t = [t, t+\delta t]$. These hulls are used to check for collisions with obstacles or other agents, and subsequently compute the trajectories of agents $\trajectory{i}$, where agents stop moving after collisions. Thus, $\trajectory{i}$ captures how other agents or obstacles might be influencing the location, velocity and collision state of agent $i$.
}
\label{fig:Illustrated_scenario}
\end{figure*}

The FeAR metric is based on the intuition that agents that reduce the feasible action space of another agent have a degree of causal responsibility for the trajectory of the latter. Consider a robot cutting across the path of a pedestrian and the pedestrian slowing down to make room for the robot (as in \cref{fig:Illustrated_scenario}). The action of the robot reduces the feasible alternatives available to the pedestrian and we intuitively feel that the robot is causally responsible for the trajectory of the pedestrian.

In this section, we elaborate how we model spatial interactions~(\cref{sec:Preliminaries}) and calculate the feasible action space of agents~(\cref{sec:ComputeFeasibleActionSpace}). We use move de rigueur to account for normative expectations~(\cref{sec:MdR_definition}) while computing the FeAR metric~(\cref{sec:FeARMetric}).

\subsection{States and actions of agents
}
\label{sec:Preliminaries}
To model spatial interactions for a  time window $T$, we treat agents as entities with point-mass dynamics in two-dimensions; the orientation of agents is disregarded.
$\Agents = \{1, \ldots, k\}$  represents the set of $k$ agents interacting with each other.
$\aState = (x, y, v_x, v_y, O)$ represents the state of the world composed of the locations $(x,y)$ and velocities $(v_x, v_y)$ of all the agents, and set of static obstacles in the scene $(O)$. Here, $x=(x_1, \ldots, x_k) = (x_i)_{i \in \Agents}$, $y=(y_i)_{i \in \Agents}$, $v_x=(v_{x,i})_{i \in \Agents}$, and $v_y=(v_{y,i})_{i \in \Agents}$.

The state $\agentState{i}$ of an agent $i$  is a tuple of its location $(x_i, y_i)\in \mathbb{R}^2$, velocity $(v_{x,i}, v_{y,i})\in \mathbb{R}^2$ and collision $\chi_i(\aState) \in \{\mathrm{True}, \mathrm{False}\}$ : $\agentState{i}= (x_i, y_i,v_{x,i}, v_{y,i}, \chi_i(\aState) ) $. $\chi_i(\aState)$ is a function of the state $\aState$ as it depends on the locations obstacles and other agents, and  is $\true$ if agent $i$ is in collision with any other agent $(x_{\neg i})$ or obstacle $(o\in O)$ and $\false$ otherwise.


Actions are accelerations which are assumed to stay constant during the time window $[0,T]$ (in accordance with the previous literature on human motor control)~\cite{durraniNewCarfollowingModel2024, gawthropIntermittentControlComputational2011, markkulaSustainedSensorimotorControl2018}.
The action chosen by agent $i$ is denoted as $\action{i} = \left(|\action{i}|, \theta_{a,i}\right)$, where $|\action{i}|$ is the magnitude and $\theta_{a,i}$ is the direction of the acceleration.
We assume that all agents choose their actions simultaneously and have the same range of actions: 
$0 \leq |\action{i}| \leq a_{\mathrm{max}}$ and $\pi \leq \theta_{a,i} \leq \pi$.
Thus, the action space of agent $i$, represented as $\ActionSpace{i}$, has two orthogonal dimensions $a$ and $\theta$ corresponding to  $|\action{i}|$ and $\theta_{a,i}$ respectively. The joint action of all the agents is denoted as $\jointAction  = \left(\action{i}\right)_{i \in \Agents} $, and the joint action space of all the agents $\JointActionSpace = \times_{i=1}^{k} \ActionSpace{i}$, has $2k$ dimensions.

We represent the trajectory of agent $i$ during the time interval $[0, T]$ as $\trajectory{i}=\{\agentState{i}(t)\}_{t \in [0, T]}$, where, $\agentState{i}(t)$ is the state of agent $i$ at time $t$. 
We are interested in quantifying the causal responsibility of agent $i$ on the trajectory $\trajectory{j}(\aState,\jointAction)$ of agent $j$ for initial state $\aState$ at $t=0$ and joint action $\jointAction$ during the time interval $[0, T]$. 

Starting from $\agentState{i}(0)$ based on $\aState$, we forward simulate the constant acceleration trajectories $\widehat{\trajectory{i}}(\agentState{i}(0),\action{i}) = \left\{\widehat{\agentState{i}}(t)\right\}_{t \in [0, T]}$,  where, $\widehat{\agentState{i}} = (\widehat{x_i}, \widehat{y_i}, \widehat{v_{x,i}}, \widehat{v_{y,i}})$ which disregards collisions with other agents or obstacles. Collision checks on $\widehat{\trajectory{i}}$ are used to compute $\trajectory{i}(\aState,\jointAction)$, such that colliding agents become stationary after collisions.

To check for collisions we partition the time window $[0, T]$ into time intervals :
\begin{equation}
\begin{aligned}
\Delta T_t&= \left[ (t-1) \delta t,\  t \delta t \right], \ 
\delta t =  \frac{T}{N_t}, \ 
[0, T] = \bigcup_{t=1}^{N_t} \Delta T_t.
\end{aligned}
\end{equation}
\emph{Trajectory hulls}, which are convex hulls of the start and end locations of the agents for each time interval (as shown in \cref{fig:Illustrated_scenario}), are used to check for collisions within a time interval.

Trajectories without collisions are not affected by other agents or obstacles i.e., $\agentState{i}(t) = \left(\widehat{\agentState{i}(t)}, \chi_i=\false \right)$ for $t \in [0, T]$, and are considered to be feasible. Thereby, we use these collision checks to model how spatially interacting agents influence the feasibility of each other's trajectories.

\begin{figure*}[t]
	\centering
	\hfill
	\begin{subfigure}[c]{0.4\linewidth}
		\begin{subfigure}[c]{0.485\linewidth}
			\centering
			\includegraphics[width=\linewidth]{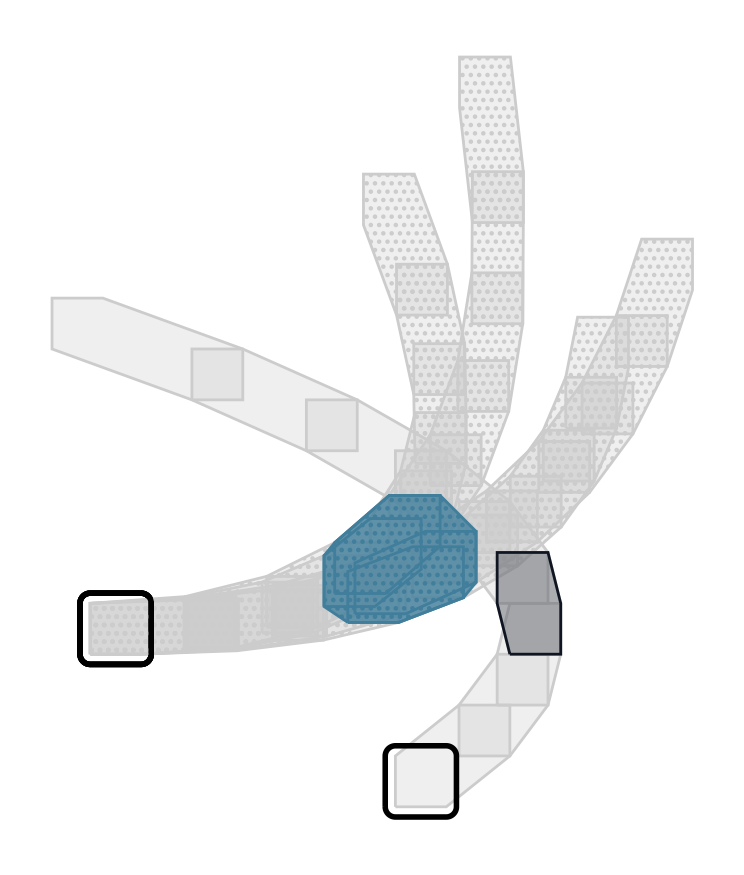}
			\captionsetup{justification=centering}
			\caption*{$\Delta T_4$: no collision}
		\end{subfigure}
		\hfill
		\centering
		\begin{subfigure}[c]{0.485\linewidth}
			\centering
			\includegraphics[width=\linewidth]{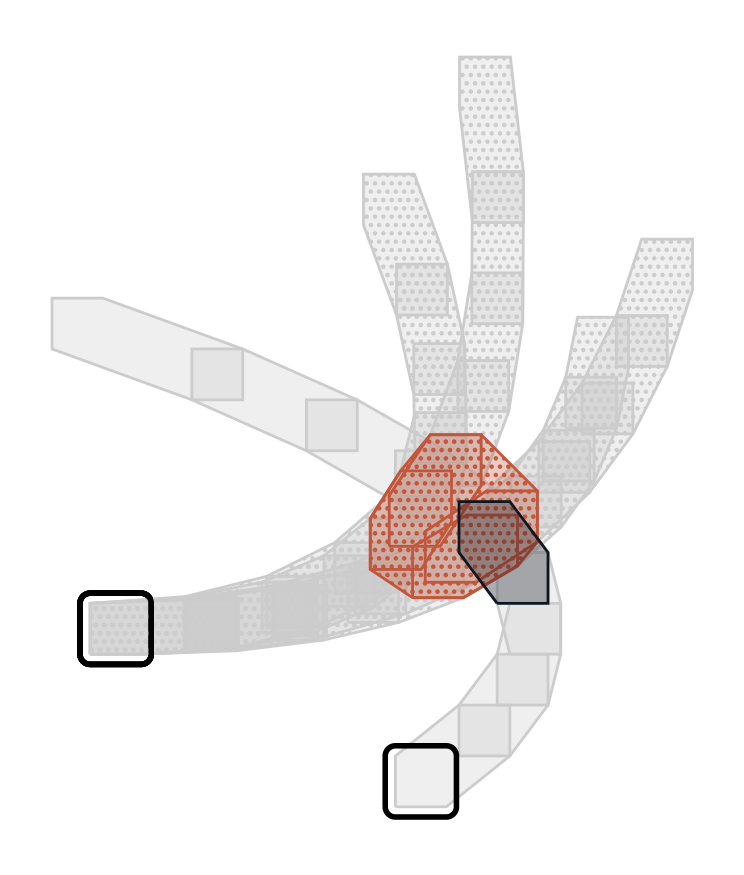}
			\captionsetup{justification=centering}
			\caption*{$\Delta T_5$: collision}
		\end{subfigure}
		\vspace{-3pt}
        \caption{Collision checks time intervals.}
		\label{fig:ActionSubspaceCollisionCheck}
	\end{subfigure}
	\hfill
	\begin{subfigure}[c]{0.55\linewidth}
	\begin{subfigure}[c]{0.3\linewidth}
		\centering
		\includegraphics[width=\textwidth]{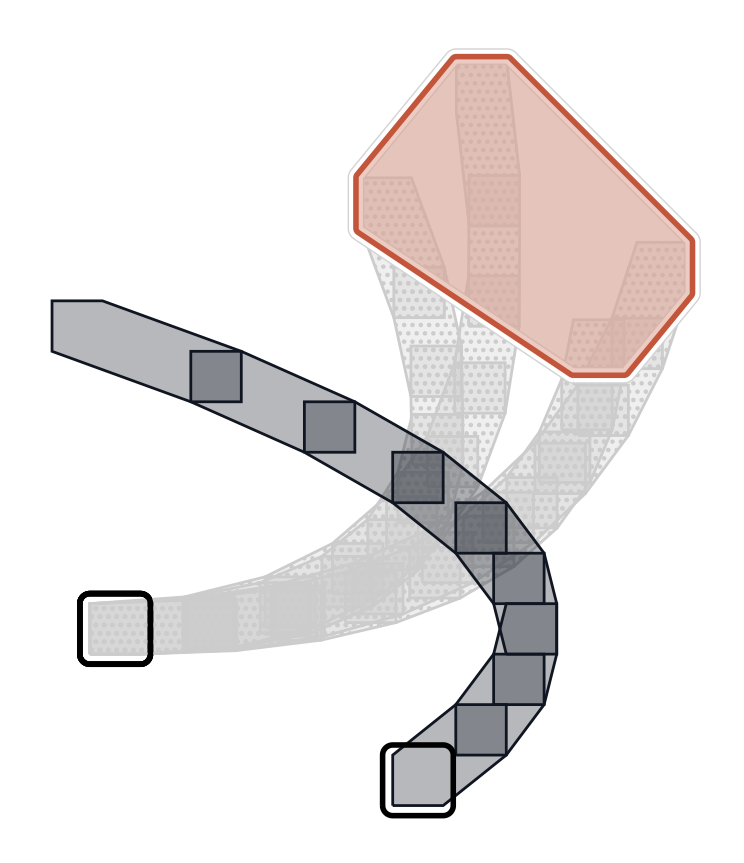}
	\end{subfigure}
	\hfill
	\begin{subfigure}[c]{0.73\linewidth}
		\centering
		\includegraphics[width=\textwidth]{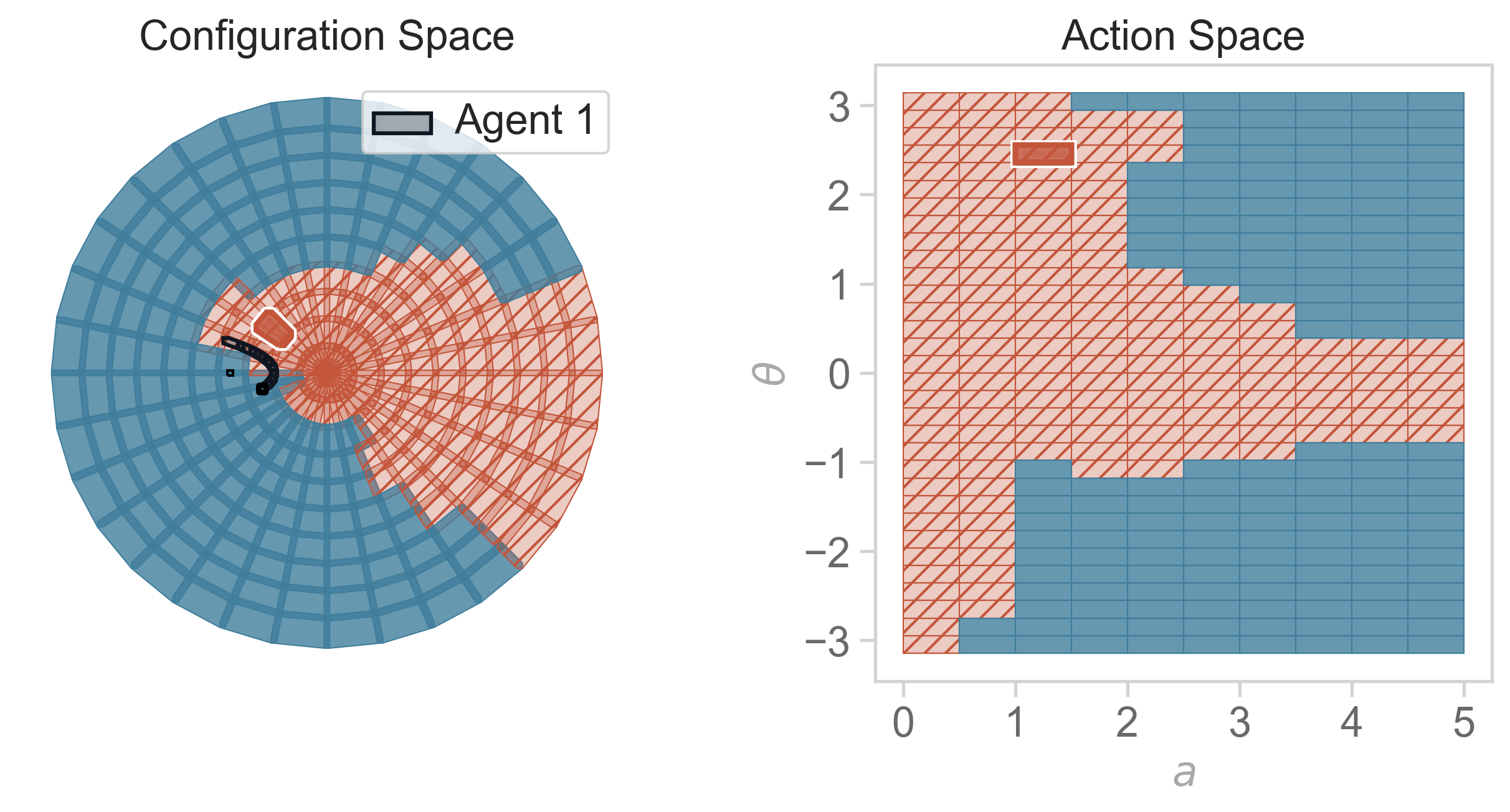}
	\end{subfigure}
	\vspace{3pt}
    \caption{Feasibility of final configurations and the feasible action space.}
	\label{fig:ActionAndConfigSpaces}
	\end{subfigure}
	\caption{\textbf{Computing the Feasible Action Space for one agent.} We divide the action space of an agent into subsets and mark an action subset as feasible (blue) if all the actions within that subset lead to feasible trajectories. Two consecutive time intervals without and with collisions are shown in (a). Action subsets leading to collisions are marked infeasible (red) in the final configuration space and action space plots (b). }
	\label{fig:subsets}
\end{figure*}

\subsection{Feasible action space and its hypervolume}
\label{sec:ComputeFeasibleActionSpace}

The cornerstone of our FeAR metric is the concept of \textit{feasible action space}. In this paper, we formalize it based on the intuition that actions leading to collisions with other agents or obstacles are to be avoided, and are therefore considered \textit{in}feasible. Therefore, for agent $j$, its feasible action space $\feasibleActionSpace{j}(\aState, \jointAction) \subset \ActionSpace{j}$ contains all actions of that agent that do not lead to collisions with obstacles or other agents (given certain actions of other agents)\footnote{We use this simple definition in the rest of the paper, but our FeAR metric does not prescribe any specific definition of feasible action space. It is compatible with alternative conceptualizations of feasible action space which could include, for instance, actions that do not lead to near-collisions or violations of traffic rules.}.

To quantify how an action of an agent $i$ affects the feasible action space of another agent $j$, we need to measure the hypervolume of the feasible action space of $j$ ($\hypervolume_{j} = \hypervolume(\feasibleActionSpace{j})$). To this end, we (1) discretize the action space into subsets, (2) determine which subsets are feasible, and (3) determine the total hypervolume of feasible action subsets.

First, we define a partition $\psi_j$ of the action space $\ActionSpace{j}$ of agent $j$  into subsets by uniformly and independently partitioning the ranges of the magnitude $|\action{j}|$ and direction $\theta_{a,j}$ of the agent's action as :
\begin{equation}
\begin{aligned}
\Delta M_m &= \left[ (m-1) \deltaMagA,\  m \deltaMagA \right], 
\end{aligned}
\end{equation}
for the magnitude, and
\begin{equation}
\begin{aligned}
\Delta D_d &= \left[-\pi+ (d-1) \deltaDirA,\  -\pi + d \deltaDirA \right], 
\end{aligned}
\end{equation}
for the direction, where,
\begin{equation}
\deltaMagA =  \frac{a_{\max}}{N_m}
\quad \text{and} \quad
\deltaDirA =  \frac{2 \pi}{N_d}
\end{equation}
such that,
\begin{equation}
\begin{aligned}
\left[0, a_{\max}\right] &= \bigcup_{m=1}^{N_m} \Delta M_m,\quad
\left[-\pi, \pi\right] = \bigcup_{m=1}^{N_d} \Delta D_d.
\end{aligned}
\end{equation}

A subset of the action space $\actionSubset{j} \in \psi_j$ is then defined as:
\begin{equation}
    \begin{aligned}
        \actionSubset{j} &= \{ \action{j} \in \ActionSpace{j},\  |\action{j}| \in \Delta M_m,\  \theta_{a,j} \in \Delta D_d \},\\
        \ActionSpace{j} &= \bigcup_{\actionSubset{j} \in \psi_j} \actionSubset{j}.
    \end{aligned}
\end{equation}

Second, for each subset $\actionSubset{j}$ we determine whether that subset is feasible $ \phi_j(\aState,\jointAction, \actionSubset{j})$ 
for agent $j$ given state $\aState$ and joint action $\jointAction$.
Here, we define a subset of action space as feasible ($\phi_j=1$) if \textit{all} actions within this subset are feasible (i.e., do not lead to a collision). Because our action space is two-dimensional and its partition is orthogonal, each subset is a rectangle with four vertices. These four vertices represent four actions of an agent which in turn correspond to four trajectories of that agent~(\cref{fig:ActionSubspaceCollisionCheck}). 

We check for collisions in each time interval $\Delta T_t$, by considering the \emph{action subset hull} which is the convex hull of the trajectory hulls of the affected agent when following actions in the action subset (as shown in \cref{fig:ActionSubspaceCollisionCheck}).  


If a collision happens at any time interval, such an action subset is considered infeasible ($\phi_j(\aState,\jointAction, \actionSubset{j})=0$) and marked with red. Feasible action subsets ($ \phi_j(\aState,\jointAction, \actionSubset{j})=1$) are marked in blue in \cref{fig:ActionAndConfigSpaces}. 
Each feasible action subset contributes its hypervolume given by:
\begin{equation}
    \hypervolume(\actionSubset{j}) = \deltaMagA \cdot \deltaDirA,
\end{equation}



Third, when all subsets of the action space are marked as feasible/infeasible (\cref{fig:ActionAndConfigSpaces}), the total hypervolume of the feasible action space $\nij$ can be calculated as
\begin{equation}
    \nij = \sum_{\actionSubset{j} \in \psi_{j}} \phi_j(\aState,\jointAction, \actionSubset{j}) \cdot \hypervolume(\actionSubset{j}).
\end{equation}


Once we formulate the feasible action space $\hypervolume_{j}$ of an affected agent $j$ for the action of an actor agent, we can compare how the action $\action{i}$ of an actor agent $i$ affects $\hypervolume_{j}$ as compared to some other action $a'_{i}$. To standardise the comparisons, we introduce the notion of \emph{move de rigueur}. 

\subsection{Move de rigueur}
\label{sec:MdR_definition}

Humans rely on social, legal, and cultural norms in spatial interactions with each other \cite{siebenCollectivePhenomenaCrowds2017,parnellResilientInteractionsCyclists2024,bjorklundDriverBehaviourIntersections2005}. 
Automated agents too must
behave according to these norms to navigate effectively among humans~\cite{robolApplyingSocialNorms2016,chenSocialLearningMarkov2022}. Furthermore, these norms 
are important for evaluating the causal influence of agents~\cite{halpernCauseResponsibilityBlame2015}. 

Here, we use the notion of \textit{Move de Rigueur} (MdR)~\cite{georgeFeasibleActionSpaceReduction2023a} to represent norms in spatial interactions. MdR $\mdr{i}$ stands for the action that is 
normatively expected of a given agent $i$ in a particular context. 

The MdR of agent $i$ is defined based on the magnitude ($|\mdr{i}|$) and direction ($\theta_{\mdr{},i}$) of the acceleration. $\mdr{i}(\aState)$ is conditional on the initial states of all agents $(\agentState{i}(0) \forall i \in \Agents)$, but does not depend on the actions of other agents ($\action{\neg i})$. The joint \textit{move de rigueur} for all the agents is denoted as $\JointMdR  = \left( \mdr{i} \right)_{i \in \Agents}$. 

A simple MdR would be for agents to maintain their current speed and direction $\left(|\mdr{i}|=0,\, \mdr{i,\theta}=0\right)$, in line with Newton's first law of motion.




 
In real-life interactions, however, we expect people to slow down when others are in front of them or to move to away from others. To capture the behaviour of agents under such expectations, MdR could instead be defined based on models of social interactions (e.g., the social force model~\cite{helbingSocialForceModel1995} for pedestrian interactions or RSS~\cite{shalev-shwartzFormalModelSafe2018, shalev-shwartzVisionZeroProvable2019} for vehicular traffic).

Thus, to ground causal analysis on normative expectations, when evaluating the causal responsibility of agent $i$ on the trajectory $\trajectory{j}$ of agent $j$, we compare the action $\action{i}$ if agent $i$ against the MdR $\mdr{i}$.

\fearCalcCaseStudyFigsCircularDiffNr{Collision_Check_Illustration_2Agents_U_turn_2-a_num-2025-01-06_00-04-50_scenario_0}{0}{\textbf{Calculating FeAR:} In a given scenario, to ascertain causal responsibility, we compare the actions performed by agents (a) with the expected \textit{moves de rigueur} (MdR) of the agents (b), and calculate the FeAR values (d) based on the feasible action spaces shown in (c), by comparing the hypervolumes of the feasible action spaces $\nij$ and $\nijMdR$ as shown in (e). $\FeAR_{i,j \neq i}$ represent the causal responsibility of agent $i$ on the trajectory $\trajectory{j}$ of agent $j$. $\FeAR_{i,i}$ (the diagonal elements which are highlighted) represents how all the other agents $\neg i$ affect the causal responsibility of agent $i$ for its own trajectory $\trajectory{i}$.}

\subsection{Feasible action-space reduction (FeAR) metric}
\label{sec:FeARMetric}


Our metric mathematically expresses the intuition that the causal responsibility of agent $i$ on the trajectory $\trajectory{j}$ of agent $j$ is proportional to the reduction in the feasible action-space $\hypervolume_{j}$ of agent $j$ due to the action $\action{i}$ of agent $i$.
We compute the feasible action-space reduction of the action $\action{i}$ in relation to the MdR $\mdr{i}$ of agent $i$ which is the normally expected move of agent $i$ in a particular context.
In other words, we perform counterfactual reasoning on the alternatives available to agent $j$ in the actual condition where $i$ follows $\action{i}$ compared to the counterfactual condition where $i$ performs the MdR $\mdr{i}$.

For $\jointAction$, we define an intervention $ \jointAction_{i} \leftarrow a'_{i}$ as the joint action in which all agents follow $A$ except agent $i$ whose actual action $\action{i}$ is replaced by another action $a'_{i}$. Joint action $\jointAction$ in which $\action{i}$ is replaced with the \emph{move de rigueur} of agent $i$ would then be denoted as $\jointAction_{i} \leftarrow \mdr{i}$. We denote hypervolume of the feasible action space of agent $j$ when agent $i$ follows its MdR as $\nijMdR$; $\niiMdR$ then represents the hypervolume of feasible action space of agent $i$ when all other agents $\neg i$ are following their MdRs $\mdr{\neg i}$.

Given the MdR $\mdr{i}$ of an actor agent $i$, FeAR imposed by agent $i$ on agent $j$ is computed based on how much the actual action $\action{i}$ of $i$ reduces the feasible action space of $j$ compared to when $i$ would have followed its MdR $\mdr{i}$: 

\begin{equation}
    \FeAR_{i,j \neq i}(\aState, \jointAction, \JointMdR)
     = 
        \clip{\frac{\nijMdR-\nij}{\nijMdR + \epsilon}},
 \label{Eq:FeARij}
\end{equation}
where function $Z$ clips the values of FeAR to $\left[-1,1\right]$ to aid interpretability of FeAR values:
\begin{equation}
	Z(x) = 
	\begin{cases}
		1,& \text{if } x \geq 1\\
		x,& \text{if } -1 < x < 1\\
		-1,& \text{if } x \leq -1.
	\end{cases}
\end{equation}

We define $\FeAR_{i,i}$ as an indicator of the \emph{\textbf{Fe}asible \textbf{A}ction-space \textbf{R}emaining} for agent $i$ for the joint action $\jointAction$ as compared to when all other agents $\neg i$ would have done their MdR $\left(\jointAction_{\neg i} \leftarrow \mdr{\neg i}\right)$:

\begin{equation}
    \FeAR_{i,i}(\aState,\jointAction, \JointMdR) 
    = 
		\clip{\frac{\nii}{\niiMdR + \epsilon}}.
 \label{Eq:FeARii}
\end{equation}

Adding $0<\epsilon\ll1$ to the denominator ensures that the metric is still defined when $\nijMdR$ or $\niiMdR$ are zero.

For brevity, we will henceforth refer to $\FeAR_{i,j}(\aState,\jointAction, \JointMdR)$ as $\FeAR_{i,j}$.
If agent $i$ is following its MdR $\mdr{i}$ then, by definition, $\FeAR_{i,j}=0$ $\forall j \neq i$. 
Positive $\FeAR_{i,j \neq i}$ values means that the action $\action{i}$ offers fewer feasible alternatives for $j$ as compared to $\mdr{i}$; which implies that $i$ is causally responsible for the trajectory $\trajectory{j}$. In the limiting case, where $\FeAR_{i,j}=1$, the action $\action{i}$ ensures that $\trajectory{j}$ is infeasible for any action chosen by agent $j$ and thus, $i$ is completely causally responsible for $\trajectory{j}$. Thus, the $\FeAR_{i,j}$ values indicate the degree of causal responsibility of $i$ for $\trajectory{j}$ with the limiting case $\FeAR_{i,j}=1$ indicating complete causal responsibility.

We say that an agent $i$ is being \emph{assertive} towards agent $j$ if $\FeAR_{i,j}>0$ which signifies that agent $i$ is reducing the feasible action space of $j$. Similarly, for $\FeAR_{i,j}<0$, we say that agent $i$ is being \emph{courteous} towards $j$ as $i$ increases the feasible action space of $j$. Thus, being assertive implies an increase in causal responsibility while being courteous implies a reduction in causal responsibility.

$\FeAR_{i,i}$ is an indication of the causal responsibility of agent $i$ for its own trajectory $\trajectory{i}$ in the current joint action $\jointAction$ compared to the case where all the other agents would have been following their MdR $\left(\jointAction_{\neg i} \leftarrow \mdr{\neg i}\right)$.
$\FeAR_{i,i}=0$ means that the other agents completely restricts the feasible action space of $i$ and $i$ has no causal responsibility for its trajectory being infeasible.
$\FeAR_{i,i}=1$ means that agent $i$ has at least as many feasible actions as when the other agents follow their MdR, which indicates that $i$ is as causally responsible for its own trajectory as in the MdR case. 



In the example scenario of the robot (agent 1) cutting across the path of the pedestrian (agent 2) as shown in \cref{fig:CS0_scenario}, one would intuitively consider that the robot is behaving assertively and is causally responsible for the trajectory of the pedestrian. And since the pedestrian is slowing down and making room for the robot, one would intuitively consider the pedestrian to be courteous and  has diminished causally responsibility for the trajectory of the robot. The FeAR values $\FeAR_{1,2} = 0.3 > 0$ and $\FeAR_{2,1}=-0.1<0$ calculated with zero acceleration as the MdR $|\mdr{i}|=0$ (\cref{fig:CS0_scenario_mdr})) agree with these intuitions (\cref{fig:CS0_FeAR}). 

The hypervolumes $\hypervolume_j$ of the feasible action space marked blue in \cref{fig:CS0_FeAS} are used to compute the FeAR values as shown in \cref{fig:CS0_FeAR_calc}. The actual action $\action{1}$ of the robot (agent 1) which cuts across the path of the pedestrian (agent 2) makes the actions of agent 2 with low $a$ infeasible, while the MdR $\mdr{1}$ would have made more of the low $a$ actions feasible. 
Since $\countNi{2}$ is 30\% smaller than $\countNiMdRj{1}{2}$, $\FeAR_{1,2}=0.3$.
On the other hand, the action $\action{2}$ of agent 2 which slows down, makes more of the actions of agent 1 feasible (especially for the actions with smaller values of $a$). Since, $\countNi{1}$ is 10\% greater than $\countNiMdRj{2}{1}$, $\FeAR_{2,1}=-0.1$.

When computing $\FeAR_{i,i}$ values, since we have just one other agent in this example, the same hypervolumes $\hypervolume_j$ used above are used. Since $\countNi{1} > \countNiMdRj{2}{1}$, $\FeAR_{1,1}=1$ has the maximum value and signifies that agent 1 has as much causal responsibility on $\trajectory{1}$ as when the other agent had followed its MdR. The 30\% reduction in $\countNi{2}$ as compared to $\countNiMdRj{1}{2}$ results in $\FeAR_{2,2}=0.7$, which means that causal responsibility of agent 2 on $\trajectory{2}$ is only 70\%.

Thus, in this example, the robot (agent 1) is behaving assertively towards the pedestrian (agent 2), while the pedestrian in behaving courteously towards the robot. This results in the pedestrian having diminished causal responsibility over their own trajectory while the robot retains the causal responsibility over its own trajectory.

In summary, the FeAR metric captures how spatially interacting agents influence each other's opportunity for action: 
$\FeAR_{i,j}, i\neq j$ indicates how an action of \textit{actor} agent $i$ is causally responsible for the trajectory $\trajectory{j}$ of the \textit{affected} agent $j$; $\FeAR_{i,i}$ indicates how much causal responsibility $i$ has over its own trajectory $\trajectory{i}$. Now that we have a metric for causal responsibility of agents for trajectories we can use this to analyse spatial interactions.

\begin{figure*}
\setcounter{casestudy}{0}
	\centering
	\begin{tabular}{cc}
		\begin{subfigure}[h]{0.45\linewidth}
			\centering
			\begin{subfigure}[c]{0.65\linewidth}
			\end{subfigure}
			\hfill
			\begin{subfigure}[r]{0.3\linewidth}
			\end{subfigure}
		\end{subfigure} &
		\begin{subfigure}[h]{0.45\linewidth}
			\centering
			\begin{subfigure}[c]{0.65\linewidth}
			\end{subfigure}
			\hfill
			\begin{subfigure}[r]{0.3\linewidth}
			\end{subfigure}
		\end{subfigure} \\ 
		\multicolumn{2}{c}{\includegraphics[width=0.35\linewidth]{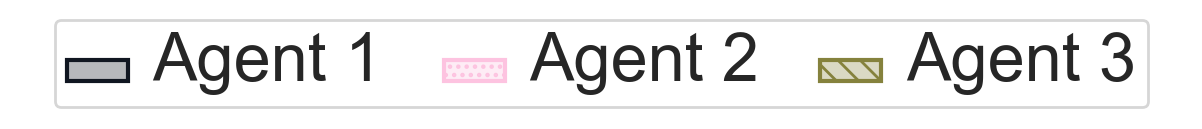}}\\
		\arrayrulecolor{black}\hline
		\arrayrulecolor{lightgray!60}\hline
		\multicolumn{2}{c}{
			\begin{tabular}{c|c}
				1. Two agents with obstructing paths & 2. Three agents with crossing paths\\ \hline
				\begin{minipage}{0.45\linewidth}
					\centering
					\CaseStudySubFigs{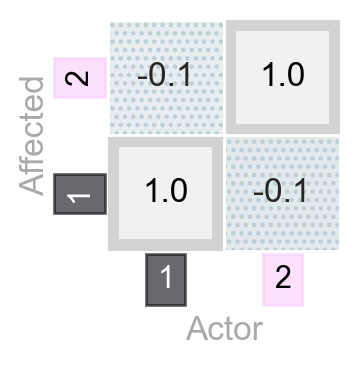}{1}{Courteous, courteous}\\
					\CaseStudySubFigs{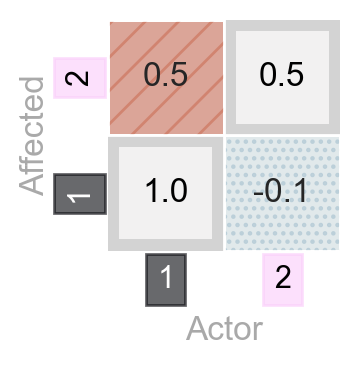}{2}{Assertive, courteous}\\
					\CaseStudySubFigs{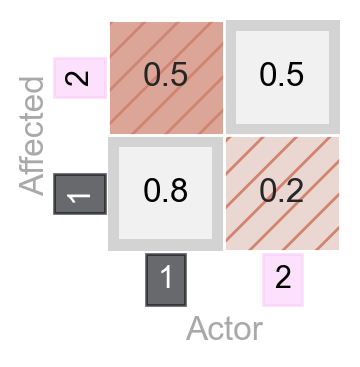}{3}{Assertive, assertive}
				\end{minipage} &
				\begin{minipage}{0.45\linewidth}
					\centering
					\CaseStudySubFigs{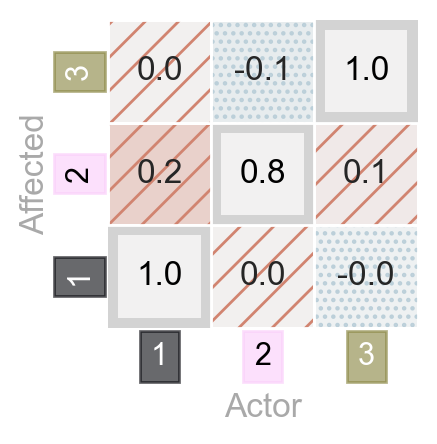}{4}{Collision between 1 and 2}\\
					\CaseStudySubFigs{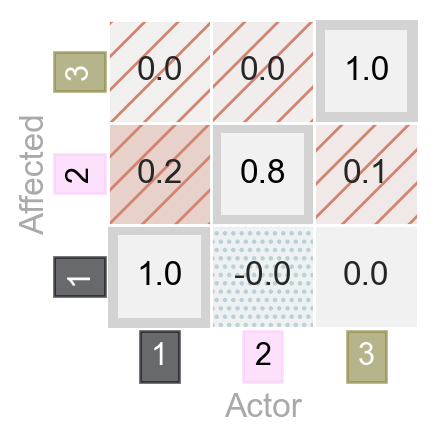}{5}{No Collisions}\\
					\CaseStudySubFigs{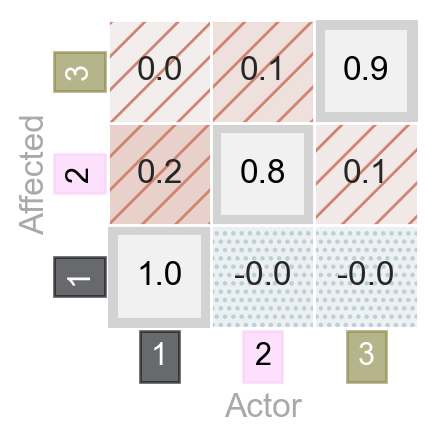}{6}{Collision between 2 and 3}
				\end{minipage} \\
				\arrayrulecolor{black}\hline
			\end{tabular}
		}\\
		\multicolumn{2}{c}{3. Effect of obstacles on FeAR} \\ \hline 
		Lane & Intersection \\ \hline
		\begin{minipage}[t]{0.45\linewidth}
			\centering
			\CaseStudySubFigs{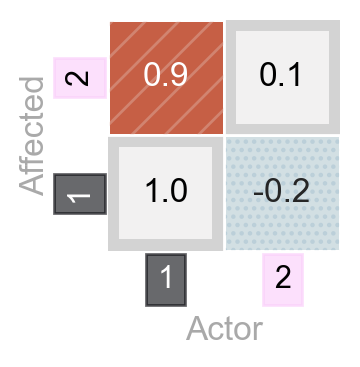}{7}{Lane - obstacles}
		\end{minipage} &
		\begin{minipage}[t]{0.45\linewidth}
			\centering
			\CaseStudySubFigs{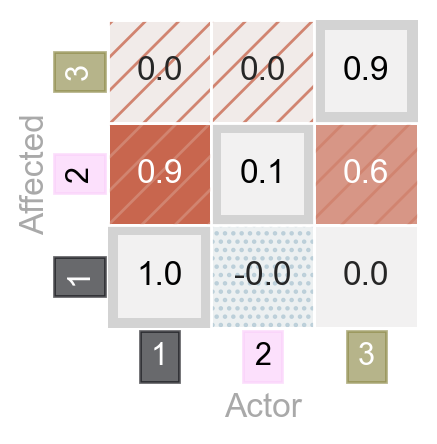}{8}{Intersection - obstacles}
		\end{minipage} \\ \arrayrulecolor{black}\hline
		\multicolumn{2}{c}{4. Counter-intuitive relation between speeding up and courteousness} \\ \arrayrulecolor{lightgray!60}\hline
		\begin{minipage}[t]{0.45\linewidth}
			\centering
			\CaseStudySubFigs{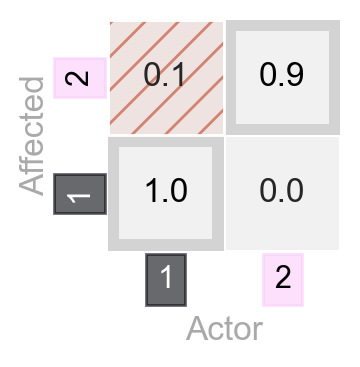}{9}{Slowing but assertive}
		\end{minipage} &
		\begin{minipage}[t]{0.45\linewidth}
			\centering
			\CaseStudySubFigs{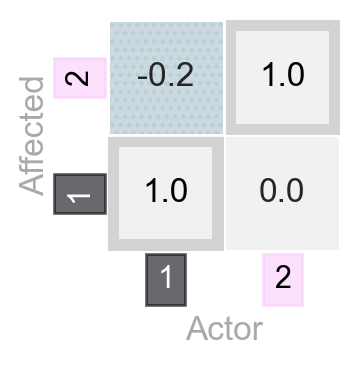}{10}{Speeding but courteous}
		\end{minipage} \\ \arrayrulecolor{black}\hline
		5. Parallel interactions without crossing paths & 6. The caveat of causal overdetermination \\ \arrayrulecolor{lightgray!60}\hline
		\begin{minipage}[t]{0.45\linewidth}
			\centering
			\CaseStudySubFigs{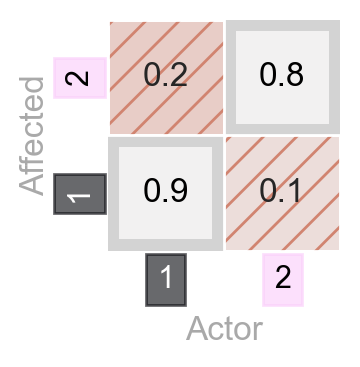}{11}{Parallel Interaction}
		\end{minipage} &
		\begin{minipage}[t]{0.45\linewidth}
			\centering
			\CaseStudySubFigs{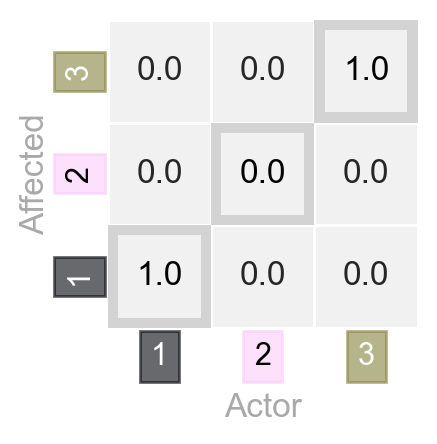}{12}{Causal Overdetermination}
		\end{minipage} \\ \arrayrulecolor{black}\hline
	\end{tabular}
	\caption{\textbf{Properties of the FeAR metric in prototypical space-sharing conflicts:} These case studies illustrate the properties of FeAR when applied to prototypical space sharing conflicts. $\FeAR_{i,j}>0$ indicates that agent $i$ is being assertive towards agent $j$, and courteous if $\FeAR_{i,j}<0$. In all these case studies we assume that the move de rigueur for all the agents is to keep moving in their current direction with the current speed (i.e., have zero acceleration).
	}
	\label{fig:CaseStudies}
\end{figure*}

\begin{table}[!htb]
    \centering
    \caption{Relation between case studies analysed in \cref{sec:Properties} and prototypical space-sharing conflicts~\cite{markkulaDefiningInteractionsConceptual2020}: obstructed paths~\textbf{OP}, merging paths~\textbf{MP}, crossing paths~\textbf{CP}, unconstrained head-on paths~\textbf{UHP}, and constrained head-on paths~\textbf{CHP}}
    \label{tab:SpaceSharingConflicts_Cases}
    \arrayrulecolor{lightgray!60} 
    \renewcommand{\arraystretch}{1.5} 
    \begin{tabularx}{\linewidth}{|c|>{\centering\arraybackslash}X>{\centering\arraybackslash}X>{\centering\arraybackslash}X>{\centering\arraybackslash}X>{\centering\arraybackslash}X|}
    \hline
        ~ & \multicolumn{5}{c|}{Space-sharing conflicts} \\ \hline
        ~Case Studies~ & \textbf{OP} &  \textbf{MP} & \textbf{CP} & \textbf{UHP} & \textbf{CHP} \\ \hline
        \textbf{\ref{fig:CaseStudy1}, \ref{fig:CaseStudy2}, \ref{fig:CaseStudy3}} & $\checkmark$ & * & ~ & ** & ~ \\ 
        \textbf{\ref{fig:CaseStudy4}, \ref{fig:CaseStudy5}, \ref{fig:CaseStudy6}} & $\checkmark$ & * & $\checkmark$ & ~ & ~ \\ 
        \textbf{\ref{fig:CaseStudy7}} & $\checkmark$ & * & ~ & ~ & ** \\ 
        \textbf{\ref{fig:CaseStudy8}} & $\checkmark$ & * & $\checkmark$ & ~ & ~ \\ 
        \textbf{\ref{fig:CaseStudy9}, \ref{fig:CaseStudy10}} & ~ & * & $\checkmark$ & ~ & ~ \\ 
        \textbf{\ref{fig:CaseStudy11}} & ~ & ~ & ~ & $\checkmark$ & ~ \\ 
        \textbf{\ref{fig:CaseStudy12}} & ~ & ~ & ~ & $\checkmark$ & ~ \\ \hline
    \end{tabularx}
    \caption*{\footnotesize * --- Because we look at just one instant of the interaction, merging paths can be considered as crossing paths at the start of the interaction and obstructed paths at the end of the interaction. \\
    ** ---  Because we disregard the orientation of agents, head-on paths can be treated as obstructed paths.}
\end{table}

\section{Properties of the FeAR metric in prototypical space-sharing conflicts}
\label{sec:Properties}

Space-sharing conflicts are central to multi-agent spatial interactions~\cite{markkulaDefiningInteractionsConceptual2020,moralesAutomatedSynthesisNormative2013}; such conflicts occur when two or more agents could potentially occupy the same space at the same time. To illustrate the applicability of the FeAR metric to spatial interactions, we analysed twelve case studies (\cref{fig:CaseStudies}, \cref{tab:SpaceSharingConflicts_Cases}) representing different combinations of five prototypical space-sharing conflicts~\cite{markkulaDefiningInteractionsConceptual2020}. In all these cases, we consider that the \textit{Move de Rigueur} (MdR) of all agents is to maintain their current speed and direction $\left(|\mdr{i}|=0,\, \mdr{i,\theta}=0\right)$. Subsequently, because normative expectations of observers also affect judgements about spatial interactions~\cite{markkulaDefiningInteractionsConceptual2020}, we investigated how different normative expectations about the actions of agents as captured by MdRs affect the causal responsibility of agents.

\subsection{Assertive and courteous actions of two agents with obstructing paths}

We start with a simple scenario where an agent (agent 1) is trailing behind another agent (agent 2) and illustrate how FeAR can be used to understand how the actions of agents affect the casual responsibility of agents for each other's trajectories.

First, we consider the case where agent 1 following agent 2 slows down while agent 2 accelerates away from agent 1 (\cref{fig:CaseStudy1}).
Compared to the MdR of moving with constant velocity, the actions of both agents increase the feasible action space of each other; resulting in negative FeAR values ($\FeAR_{1,2}=-0.1$ and $\FeAR_{2,1}=-0.1$). The negative FeAR values imply that the agents are behaving courteously towards each other and are not causally responsible for each other's trajectory. As a consequence, both agents have the maximum causal responsibility for their own trajectories as reflected in $\FeAR_{i,i}$ values ($\FeAR_{1,1}=1, \, \FeAR_{2,2}=1$).

In the second case study, we consider a variation of the previous case (\cref{fig:CaseStudy1}) where agent 1 is accelerating in the direction of agent 2 while agent 2 still accelerates away from agent 1 (\cref{fig:CaseStudy2}). The acceleration of agent 1 halves the feasible actions available to agent 2 as it cannot slow down now. $\FeAR_{1,2}=0.5$ is positive indicating that agent 1 is acting assertively towards agent 2 and $\FeAR_{2,2}=0.5 < 1$ is indicative of the diminished causal responsibility of agent 2 over its own trajectory. Thus, by accelerating, agent 1 is 50\% causally responsible for the trajectory of agent 2.
$\FeAR_{2,1}=-0.1$ and $\FeAR_{1,1}=1$ remain the same as the last case because agent 2 is doing the same action as before; and indicates that agent 2 is behaving courteously towards agent 1.

Third, we consider the case where agent 1 is accelerating towards agent 2 (same as the previous case)
and agent 2 is lightly slowing down  
in front of agent 1 (\cref{fig:CaseStudy3}); 
the magnitude of acceleration of agent 1 is three times the magnitude of deceleration of agent 2 ($\left| a_1 \right| = 3 \left| a_2 \right|$). Just as in the previous case,
$\FeAR_{1,2}=0.5>0$ indicates that agent 1 is acting assertively towards agent 2. By slowing down in front of agent 1, agent 2 restricts the feasible action space of agent 1 and restricts it from accelerating. $\FeAR_{2,1}=0.2>0$ indicates that, agent 2 is acting assertively towards agent 1. Thus, both agents share some causal responsibility for each other's trajectory. Both $\FeAR_{1,1}$ and $\FeAR_{2,2}$ are less than 1, indicating that both agents have diminished agency to avoid a collision and are less causally responsible for their own collision. Furthermore, in this case study, agents 1 and 2 collide as a result of their chosen actions. Considering agent 1 as the affected agent, $\FeAR_{1,1}=0.8$ and $\FeAR_{2,1}=0.2$ indicate that agent 1 has greater causal responsibility for its collision than agent 2. From the perspective of agent 2, $\FeAR_{2,1}=0.5$ and $\FeAR_{2,2}=0.5$, indicating that both agents 1 and 2 have the same causal responsibility for the collision of agent 2. Since, FeAR treats the collision of each agent separately, it can be useful in distributing the costs of damages to individual vehicles, which might not be uniform across all colliding vehicles.

Together, the three variations of the scenario with two agents on obstructed paths illustrate how positive values of $\FeAR_{i,j}$ are indicative of the degree of causal responsibility of agent $i$ on the trajectory $\trajectory{j}$ of agent $j$ when agent $i$ is behaving assertively towards agent $j$.
Negative FeAR values show when agents are behaving courteously, which is to be encouraged when agents are spatially interacting with each other.
Furthermore, instead of treating accelerations or decelerations as good or bad on their own right, FeAR provides a contextual analysis of the impact of these actions in relation to different actors and their states. 

\subsection{Multi-agent interaction with obstructed and crossing paths}

Next, we consider three variations of a case where agent 2 is caught between agent 1 (approaching agent 2 from behind) and agent 3 (cutting across the path of agent 2 in front of it); both agents 1 and 3 are starting from rest while agent 2 is already in motion. 
We consider three variations based on the action of agent 2 --- one where agent 2 decelerates and collides with agent 1 (\cref{fig:CaseStudy4}), one where agent 2 has a low acceleration and is collision-free (\cref{fig:CaseStudy5}), and one where agent 2 accelerates fast and collides with agent 3 (\cref{fig:CaseStudy6}).

In all three cases, irrespective of the collision of agent 2, $\FeAR_{1,2}=0.2$ and $\FeAR_{3,2}=0.1$ indicate that by accelerating towards of cutting across the path of agent 2, both agents 1 and 3 are being assertive towards it.$\FeAR_{1,2}>\FeAR{3,2}$ also implies that agent 1 has more causal responsibility on $\trajectory{2}$ than agent 3.
$\FeAR_{2,2}=0.8<1$ reflects the reduction in causal responsibility of agent 2 for $\trajectory{2}$ due to the collective action of agents 1 and 3.
Because values $\FeAR_{i,2}$ do not depend on the actions of agent 2, they provide a holistic picture of the causal responsibility of other agents on the trajectory of 2. The independence of $\FeAR_{i,j}$ from the actions of the affected ($\action{j}$) is an important property of the FeAR metric.
Furthermore, these cases illustrate how the FeAR metric captures causal responsibility of agents for near-misses as in~(\cref{fig:CaseStudy5}) and for cases where one agent's nudge causes other agent to collide with a third agent~(\cref{fig:CaseStudy4,fig:CaseStudy6}).

Considering the influence of agent 2 on agents 1 and 3, the values of $\FeAR_{2,i}$ show how the same agent can be assertive to some agents while being courteous to other agents. When agent 2 is decelerates and crashes with agent 1, $\FeAR_{2,1}>0$ and $\FeAR_{2,3}<0$ show that it is being assertive to agent 1 while being courteous to agent 3~(\cref{fig:CaseStudy4}). The opposite is true when agent agent 2 is accelerating~(\cref{fig:CaseStudy5,fig:CaseStudy6});  $\FeAR_{2,1}<0$ and $\FeAR_{2,3}>0$ show that agent 2 is being assertive to agent 3 while being courteous to agent 1. The magnitude of the FeAR values also indicate that agent 2 has more causal influence on agent 1 as compared to agent 3. So, given a choice between the actions of agent 2 in the first~(\cref{fig:CaseStudy4}) and last~(\cref{fig:CaseStudy4}) cases, the socially optimal one would be accelerating where agent 2 is courteous to agent 1 ($\FeAR_{2,1}=-0.1$) and only lightly assertive to agent 3 ($0>\FeAR_{2,3}>>0.1$); so that the causal responsibility of agent 2 on other agents is minimised. 


These cases demonstrate how the FeAR metric can be used to understand how multiple agents can be causally responsible for one agent and how one agent can be assertive to some agents while simultaneously being courteous to others. Additionally, the magnitudes of the $\FeAR_{i,j}$ values are useful for comparing the degrees of causal responsibility between different pairs of agents with either the same actor ($i$) or affected ($j$) agent.


\subsection{Obstacles for modelling constrained paths}

In the cases presented so far, agents were moving in an open space without restrictions from the environment. In real life however, static obstacles and constraints like road lanes can further limit the feasible actions available to agents. To illustrate the influence of such restrictions on FeAR, we consider two cases: an obstructed-path interaction between two agents in a corridor (\cref{fig:CaseStudy7}, derived from case (b)) and an interaction between three agents at an intersection with lane boundaries (\cref{fig:CaseStudy8}, derived from case (e)).

In both cases, adding static obstacles reduces the feasible action space for both the MdR actions and the actual actions. This decreases the denominator in \cref{Eq:FeARij}  and thus the corresponding $\FeAR_{i,j \neq i}$ values are higher than without the static obstacles (cf. \cref{fig:CaseStudy2} and \cref{fig:CaseStudy7}, as well as \cref{fig:CaseStudy5} and \cref{fig:CaseStudy8}). Intuitively, when an agent has fewer feasible actions to chose from, the impact of reducing some of those actions is amplified. More formally, when evaluating $\FeAR_{i,j}$, restrictions imposed on $j$ by obstacles or other agents ($\neg i$) will amplify the values of $\FeAR_{i,j \neq i}$. Thus, if an agent further restricts an already quite restricted agent, the FeAR value will be higher than in a hypothetical case when the same action restricts an agent with a larger initial feasible action space. On the other hand, if an agent is courteous to an otherwise restricted agent, it will have large negative value of FeAR, expressing that the affected agent's actions space is substantially expanded.

We are not making any claims on the impact of obstacles on $\FeAR_{i,i}$ as it depends on the actions of all other agents $\neg i$, and some of those might be assertive while some might be courteous; which makes it uncertain whether $\FeAR_{i,i}$ will increase, decrease or stay the same.

To summarise, restrictions in the environment reduce the affected agent's feasible action space for the MdR and thereby amplify $\FeAR_{i,j \neq i}$ values. 

\subsection{Counter-intuitive relation between speeding up and courteousness}

Intuitively, we associate assertive behaviour with moving fast; for instance, humans tend to blame those who speed up for negatively influencing others around them \cite{alonsoSpeedRoadAccidents2015, ohernKaahaajatFinnishAttitudes2023}. Here, we use analyses based on FeAR values to highlight two cases that run counter to this intuition.

We consider two crossing paths interactions between two agents already moving at some speed: one where agent 1 is decelerating~(\cref{fig:CaseStudy9}) and one where agent 1 is accelerating~(\cref{fig:CaseStudy10}). 

According to the FeAR metric, when agent 1 slows down, it is being assertive towards agent 2 ($\FeAR_{1,2}=0.1$ is positive, \cref{fig:CaseStudy9}), and when it accelerates, it is being courteous ($\FeAR_{1,2}=-0.2$ is negative, \cref{fig:CaseStudy10}). This happens because, when agent 1 accelerates, it quickly moves out of the way of agent 2 and when slowing down, it blocks more actions of agent 2. Thus, in this scenario, by slowing down, agent 1 is causally responsible for $\trajectory{2}$.

Hence, in addition to the presence of other agents, FeAR takes into account the dynamic context including velocities and accelerations, allowing for more nuanced measurement of causal responsibility.

\subsection{Parallel interaction without intersecting paths}
 
To highlight the applicability of the FeAR metric in spatial interactions that are beyond the scope of traditional analysis approaches and metrics (such as time-to-collision (TTC) and post-encroachment time (PET)), we consider a case derived from the prototypical ``unconstrained head-on paths'' interaction \cite{markkulaDefiningInteractionsConceptual2020}. Here, two agents are accelerating past each other but their paths do not intersect (\cref{fig:CaseStudy11}). In this scenario, there is formally no space-sharing conflict, and traditional interaction metrics such as TTC and PET are undefined. However, agents still might take actions (for instance, to avoid passing too close to each other), potentially affecting each other, which makes this case relevant for quantifying responsibility.

The positive values of FeAR $\FeAR_{1,2}=0.2 >0$ and $\FeAR_{2,1}=0.1 >0$ show that agents 1 and 2 are assertive to each other even though their paths do not cross. Even though these actions do not lead to collisions, their actions reduce the  number of feasible actions available for the other agent. Furthermore, as agent 1 has a higher acceleration than agent 2, $\FeAR_{1,2} > \FeAR_{2,1}$. This implies that agent 1 is more assertive and has greater causal responsibility for $\trajectory{2}$ than the other way around.

Thus, the FeAR metric can be applied to understand how agents influence each other even when they are not heading to a collision or have crossing paths.

\subsection{The caveat of causal overdeterminination}
Causal overdetermination happens when more than one agent is completely causally responsible for an outcome, for example when two people simultaneously throw rocks at a bottle and the bottle breaks \cite{schafferOverdeterminingCauses2003}. Consider a case where agents 1 and 3 are both accelerating from standstill towards agent 2 which is stationary between them(\cref{fig:CaseStudy12}). In this case, $\FeAR_{2,2}=0$, which indicates that agent 2 has no causal responsibility over $\trajectory{2}$ because it has no feasible actions left under the current actions of agents 1 and 3. At the same time, it would have some feasible moves if all the other agents had followed their MdR. So collectively agents 1 and 3 have an influence on agent 2, but these agent's individual FeAR values ($\FeAR_{1,2}=0$ and $\FeAR_{3,2}=0$) indicate that each of them individually does not bear any responsibility for restricting agent~2. 

This seemingly paradoxical situation happens because even when either agent 1 or 3 is following the MdR, the other agent would still be colliding with agent 2. In this case, the collision of agent 2 is causally overdetermined by the actions of both agents 1 and 3. 
Our definition of the FeAR metric is aimed to capture the causal responsibility of actions of individuals and does not extend to 
causal influences of groups (such as in cases of causal overdetermination); for such cases, group responsibility~\cite{loriniLogicalAnalysisResponsibility2014, gladyshevGroupResponsibilityExceeding2023, yazdanpanahDistantGroupResponsibility2016, yazdanpanahApplyingStrategicReasoning2021, alechina2017causality} should be considered.

\begin{figure*}
    \centering
    \begin{subfigure}[c]{0.9\linewidth} 
    \begin{subfigure}[c]{0.3\linewidth}
    \caption*{}
    \end{subfigure}
    \begin{subfigure}[c]{0.6\linewidth}
    \captionsetup{justification=centering}
    \caption*{$\FeAR_{i,j}$ values for different MdRs}
    \hrulefill
    \end{subfigure}
    \begin{subfigure}[c]{0.35\linewidth}
    \caption*{Scenario}
    \end{subfigure}
    \hfill
    \begin{subfigure}[c]{0.2\linewidth}
    \captionsetup{justification=centering}
    \caption*{$\JointMdR=\Compliant$}
    \end{subfigure}
    \hfill
    \begin{subfigure}[c]{0.2\linewidth}
    \captionsetup{justification=centering}
    \caption*{$\JointMdR=\Assertive$}
    \end{subfigure}
    \hfill
    \begin{subfigure}[c]{0.2\linewidth}
    \captionsetup{justification=centering}
    \caption*{$\JointMdR=\CompliantLeft$}
    \end{subfigure}
    \\ \hrulefill \\
    \MdRCaseStudySubFigsBarsPicked{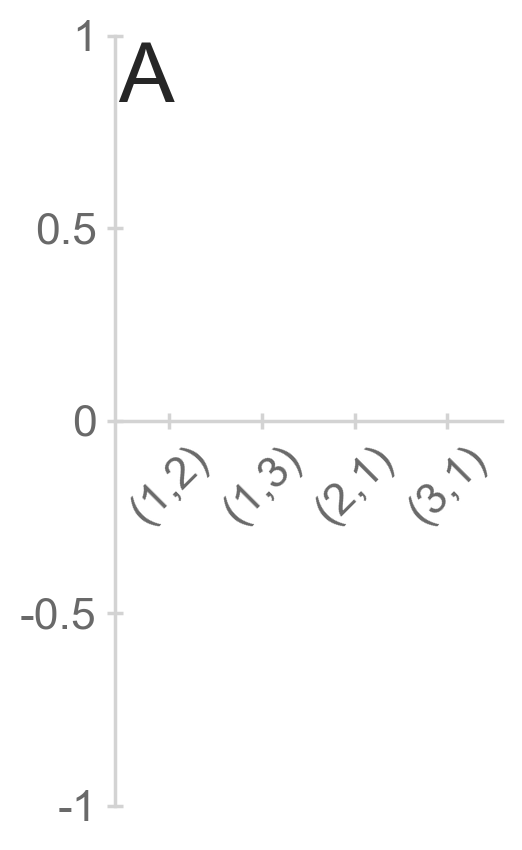}{B}{Compliant}
    \MdRCaseStudySubFigsBarsPicked{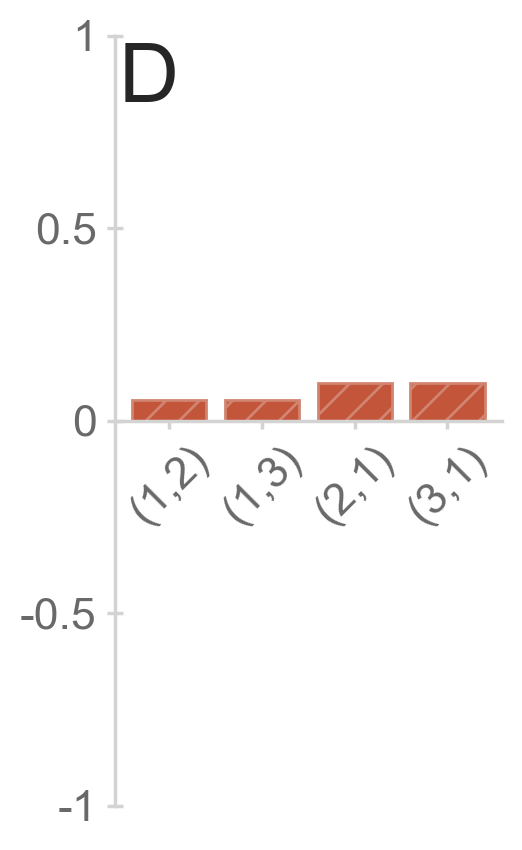}{C}{Assertive}
    \MdRCaseStudySubFigsBarsPicked{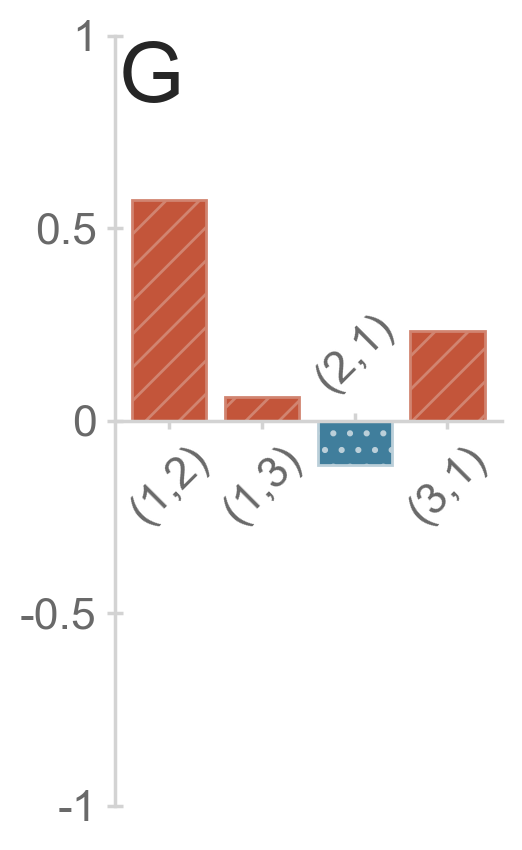}{D}{Compliant-Left}

    \end{subfigure}
    \caption{\textbf{The effect of \textit{Move de Rigueur} on FeAR:} Different contexts might warrant different expectations for the actions of agents. In the interaction of one agent going head-on towards two other agents, different joint actions $\jointAction$ (rows) and different choices of MdR $\JointMdR$ (columns) lead to different FeAR values. For brevity, only four most informative FeAR values are shown for each case. Denoting the vertical position of the agent as $y$, and the attraction to the center of the lane as $k_{\text{lane}}$, the joint actions/MdRs are defined as:
    }
    \label{fig:CaseStudiesMdRBars}
    \begin{align*}
        \Compliant&: \text{Yielding,} & & \text{ lanes } = \text{None}\\
        \Assertive&: \text{Unyielding from the centre,} & k_{\mathrm{lane}}=1, & \text{ lanes } = [y, y, y] \\
        \CompliantLeft&: \text{Yielding to the left lane,} & k_{\mathrm{lane}}=3, & \text{ lanes } = [y-20, y+20, y+20]\\
    \end{align*}
\end{figure*}

\subsection{The effect of \textit{Move de Rigueur} on FeAR}

Definitions of space-sharing conflicts are observer-dependent~\cite{markkulaDefiningInteractionsConceptual2020} and notions of causality can vary with varying expectations of observers across different contexts~\cite{halpernCauseResponsibilityBlame2015}. To account for normative expectations, the FeAR metric takes as input the set of \textit{Moves de Rigueur} (MdR) --- a consistent set of normatively expected actions for all the agents in a given context~\cite{georgeFeasibleActionSpaceReduction2023a}. Here, we showcase methods for defining MdRs and how the choice of MdR can affect FeAR values.

Consider a head-on interaction between an agent and a pair of agents headed in the opposite direction.
We consider three possible joint actions for the agents  (computed based on the social force model~\cite{helbingSocialForceModel1995} where pedestrians are attracted to a lane while being repelled by other pedestrians).

\begin{itemize}
    \item In $\Compliant$, all the agents behave compliantly; agents 2 and 3 split and move aside for agent 1, and agent 1 slows down when approaching agents 2 and 3 ~(\cref{fig:CaseStudyMdRB}).
    \item In $\Assertive$, the agents behave more assertively and deviate less from their original heading and speed~(\cref{fig:CaseStudyMdRC}).
    \item In $\CompliantLeft$, agents 2 and 3 move as group to their left and agent 1 swerves to its left~(\cref{fig:CaseStudyMdRD}).
\end{itemize}

We used each of $\Compliant$, $\Assertive$, $\CompliantLeft$ as the MdR, and analysed how the FeAR values for the joint actions vary under different choices of MdR (\cref{fig:CaseStudiesMdRBars}). Since all the cases here have the same state $\aState$, we use $\fearAM{i}{j}{\jointAction}{\JointMdR}$ to denote the value of $\FeAR_{i,j}(\aState, \jointAction, \JointMdR)$. We are interested in how agent 1 influences agents 2 and 3 and how agents 2 and 3 influence agent 1; and hence will only focus on $\FeAR_{1,2}$, $\FeAR_{1,3}$, $\FeAR_{2,1}$ and $\FeAR_{3,1}$. 

Trivially, when agents take the MdR as their action, the FeAR values $\fearAM{i}{j \neq i}{\JointMdR}{\JointMdR}$ are zero, (panels \textbf{A}, \textbf{E} and \textbf{I} on the diagonal of \cref{fig:CaseStudiesMdRBars}).


We start by analysing FeAR values for $\jointAction = \Compliant$ (panels \textbf{B} and \textbf{C}). When agents are expected to be unyielding ($\mdr=\Assertive$) but yield, they provide each other with additional feasible action space and behave courteously (highlighted by negative values of $\fearAM{i}{j}{\Compliant}{\Assertive}<0$ in panel \textbf{B}).

Next, we consider the norm that agents yield left $\mdr = \CompliantLeft$ (panel \textbf{C}).
Agent 1 by slowing down without changing direction, increases the feasible action space of agent 2 to its left while reducing the feasible action space of agent 3 to its right, thereby, being courteous to agent 2 ($\fearAM{1}{2}{\Compliant}{\CompliantLeft}<0$) and assertive to agent 3 ($\fearAM{1}{3}{\Compliant}{\CompliantLeft}>0$).

Agent 2 swerving right as in $\jointAction=\Compliant$ would remove the feasible actions of agent 1 that would have been available if agent 2 swerved left as in $\JointMdR=\CompliantLeft$; hence, agent 2 is being assertive to agent 1 and is causally responsible for $\trajectory{1}$ ($\fearAM{2}{1}{\Compliant}{\CompliantLeft}>0$).

Meanwhile, even though agent 3 swerves left, it does not swerve as much as in $\CompliantLeft$ and reduces the feasible action space of agent 1 and is assertive towards agent 1 ($\fearAM{3}{1}{\Compliant}{\CompliantLeft}>0$).

Moving on to the second joint action $\jointAction=\Assertive$ (panels \textbf{D} and \textbf{F}), $\fearAM{i}{j}{\Assertive}{\Compliant}>0$ shows that when agents are expected to yield, but do not, they are behaving more assertively and are more casually responsible because they reduce the feasible action space of other agents (Panel \textbf{D}).

The values of $\fearAM{i}{j}{\Assertive}{\CompliantLeft}$ (Panel \textbf{F}) are similar to $\fearAM{i}{j}{\Compliant}{\CompliantLeft}$ (Panel \textbf{C}) with a small increase in the positive values reflecting the increase in causal responsibility due to assertive actions.

Finally, for the joint action $
\jointAction = \CompliantLeft$ (panels \textbf{G} and \textbf{H}), the FeAR values for MdRs $\mdr = \Compliant$ and $\mdr = \Assertive$ have similar signs and only vary in the magnitudes. 
As compared to slowing down as in $\Assertive$ or $\Compliant$, Agent 1 by swerving left restricts more actions of agent 2 that swerve right; and is thus assertive towards agent 2 ($\FeAR_{1,2}>0$). 

By going straight as in the MdRs, agent 1 collides with and stops agent 2 midway, and prevents the swerving left motion of agent 2 from restricting more actions of agent 3. Compared to this, by swerving left, and letting agent 2 restrict agent 3,  agent 1 is being assertive to agent 3~($\FeAR_{1,3}>0$) and is more causally responsible for $\trajectory{3}$.


Agent 2 by swerving left increases the feasible action space of agent 1 and is acting courteously, more so for $\JointMdR=\Assertive$: $\fearAM{2}{1}{\CompliantLeft}{\Assertive} < \fearAM{2}{1}{\CompliantLeft}{\Compliant} < 0$.

The actions of agent 3 in the MdRs (which are also swerving left, but not as sharp as in $\CompliantLeft$), will lead to collisions with agent 2 which is swerving left. This stops agents 2 and 3 midway and increases the feasible actions for agent 1 in the MdR condition. Thus, agent 3 is acting assertively towards agent 1 in $\Assertive$ and $\Compliant$, and is causal responsible for $\trajectory{1}$; more so for $\JointMdR=\Assertive$ 
: $\fearAM{3}{1}{\CompliantLeft}{\Assertive} > \fearAM{3}{1}{\CompliantLeft}{\Compliant} > 0$.

These examples illustrate how the FeAR metric can flexibly account for different normative expectations via different \emph{moves de rigueur} and the importance of choosing the right MdR for the context in question.

\section{Use cases for the FeAR metric}
\label{sec:Applications}

\backwardResponsibilityFigs{1}{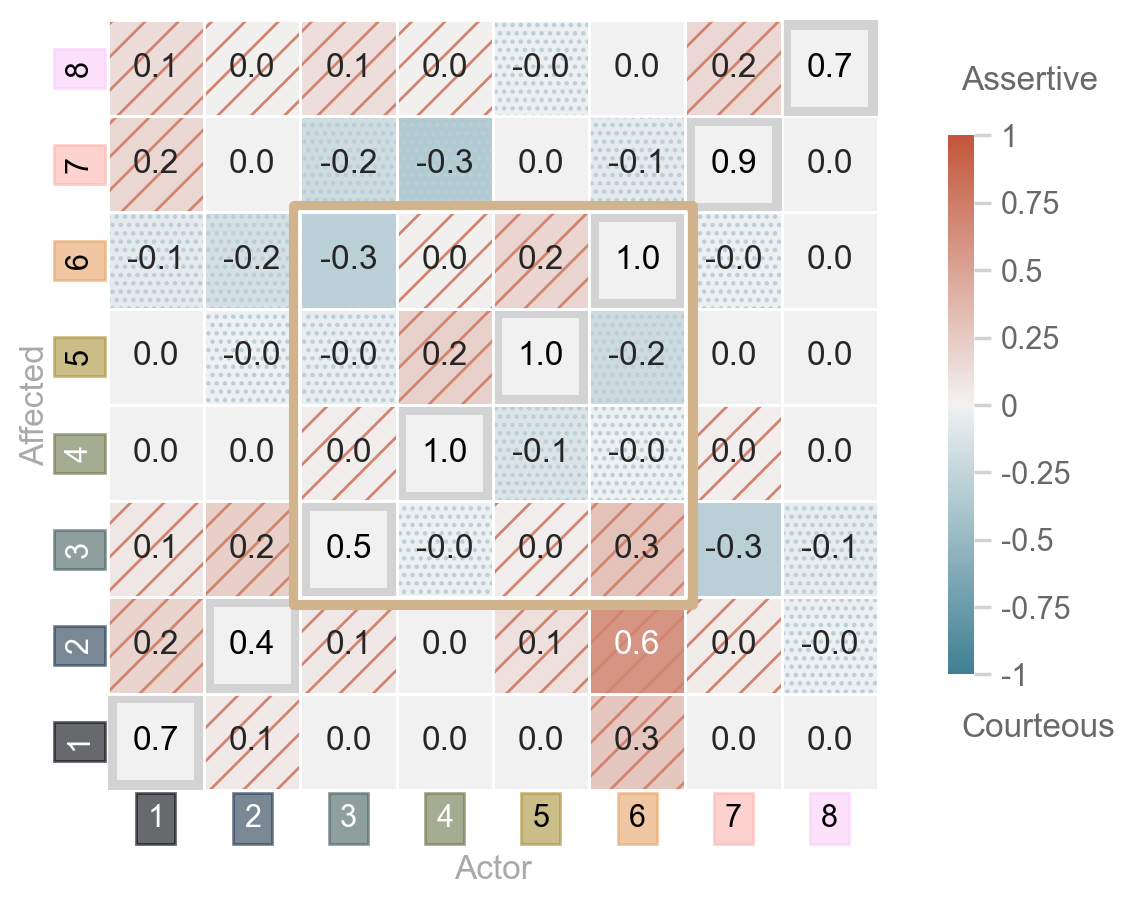}{\textbf{Using FeAR to analyse backward-looking responsibility.} (a) Eight agents interacting with each other.
At the core of the interaction, agents \textbf{3, 4, 5} and \textbf{6} take actions that bring them into a ``cyclical'' relationship. Agent 1 and 2 are moving in parallel with agent 3 (as if part of a group) and agents 7 and 8 are approaching the interaction area one behind another, but are slowing down; (b) \textit{Moves de Rigueur} for each agent, computed based the social force model to move away from each other; (c) FeAR value, with the square around the FeAR values corresponding to interactions between agents 3, 4, 5, and 6; (d) Largest FeAR values visualized as a directed graph between the agents. 
}

In this section, we illustrate two main use cases we envision for the FeAR metric: analysing sophisticated multi-agent scenarios to better understand the interactions (\cref{sec:BackwardResponsibility}) and supporting agents' decision making (\cref{sec:FowardResponsibility}).

\subsection{Analysing backward-looking responsibility}
\label{sec:BackwardResponsibility}

To illustrate the application of FeAR for understanding responsibility in a backward-looking sense (``\textit{who was responsible?}''), we used a hypothetical scenario with an entangled, gridlock-like interaction between eight agents (\cref{fig:BackwardFear_1}). 
At the centre of the interaction are four agents (3, 4, 5 and 6) that are moving past each other (\cref{fig:BackwardFear_1_fear_graph}). If one of these agents decides to stay, it will cause the agent approaching from its right to collide into it. Thus, to avoid collisions, agent 4 has to move out of the way of agent 3, agent 5 has to move out of the way of  agent 4, agent 6 has to move out of the way of agent 5 and agent 3 has to move out of the way of agent 6. If they decide to follow overly conservative policies (staying or moving with a lower speed), they will be stuck in a gridlock. Locations and actions of agents 1, 2, 7 and 8 further increase the complexity of the interaction. Agents 1 and 2 are moving in parallel with agent 3 -- this allowed us to understand the effect of agents travelling in ``herds''. Agents 7 and 8 are moving one after another head-on against agent 3 and slow down before crossing the path of agent 4.

As MdRs for each agent, we used actions computed with a social force model~\cite{helbingSocialForceModel1995, rudenkoHumanMotionTrajectory2020} which captures the natural tendency of human pedestrians to avoid collisions with others \cref{fig:BackwardFear_1_mdr}. 
While the social force model is not inherently collision-free, in our specific scenario if all agents follow these trajectories, there would not be any collisions. Importantly, this is just one way of defining a \emph{Move de Rigueur} which assumes that agents are expected to avoid collisions. Depending on the scenario, more advanced ways of defining MdR could also include considerations like traffic rules, physical constraints, and cognitive capacities of humans. 

In our scenario, agents 3, 4, 5, and 6 are a) accelerating and b) crossing the future paths of other agents, hence a naive interpretation of the interaction would implicate them as assertive and responsible for potential collisions. However, the FeAR metric suggests this is not the case: while each of these agents is being assertive to the agent in front (e.g., $\FeAR_{4,5}>0$), they are simultaneously being being courteous towards the agent on their right-hand side (e.g., $\FeAR_{4,3}<0$). It is exactly this cyclical courteousness that contributes to avoiding a gridlock.  

Similarly, there are other agents that are moving out of the way of another agent thereby giving them more feasible actions to choose from. This is seen in the how agent 4 interacts with agent 7 ($\FeAR_{4,7}<0$). On the other hand, agent 7 is slowing down in front of 8 leads to $\FeAR_{7,8}$ being positive, thus being assertive towards agent 8. 

Considering the group of three agents in the top left corner, FeAR suggests that agents 1, 2, and 3 are courteous towards agent 6 ($\FeAR_{3,6} < \FeAR_{2,6} < \FeAR_{1,6} <0 $) while only agent 1 is being being assertive towards agent 7 ($\FeAR_{1,7}>0$, $\FeAR_{2,7}=0$ and $\FeAR_{3,7}<0$). Comparing the values of $\FeAR_{i,6}$ and $\FeAR_{i,7}$, we can see that agent 3 is the one with the lowest FeAR value and agent 1 is the one with the highest which signals that agent 3 which is closest to agents 6 and 7 is more courteous to them than agent 1 following the same action as agent 3. This highlights how both the magnitude and sign of the FeAR values can be used to show that agent 1 has more causal responsibility on agents 6 and 7 than agent 3 whose movement is almost identical to that of agent 1.

\forwardResponsibilityFigs{1}{8Agents_RoundAbout_Intersection_compact_13-a_num-2024-10-16_01-17-14}{2}{\textbf{Using FeAR for responsibility-aware action selection.} In the scenario of  \cref{fig:BackwardFear_1}, we take the perspective of agent 3 and illustrate how this agent can use FeAR to choose the most socially responsible action (acceleration with magnitude $a$ and direction $\theta$). For simplicity, we assume that agent 3's predictions of other agents' actions are the same as their actions represented in \cref{fig:BackwardFear_1}. 
In (b) and (c), collisions are marked by a red cross, and optimal actions by a white dot. In (c), negative values indicate the number of agents that ego agent is being courteous to and a positive values indicate the number of agents it is being assertive to.
}

\subsection{Responsibility-aware action selection}
\label{sec:FowardResponsibility}
Sections~\ref{sec:Properties} and~\ref{sec:BackwardResponsibility} demonstrated properties of FeAR and its application as a backward-looking responsibility metric. To complement this perspective, this section focuses on forward-looking responsibility, highlighting the potential of FeAR as a metric that agents can use to choose future actions in a more socially responsible way. While existing decision making and motion planning approaches for autonomous agents typically include safety and efficiency as key metrics to optimize, the inclusion of responsibility-aware metrics is a promising but under-explored research area~\cite{shalev-shwartzFormalModelSafe2018,cosnerLearningResponsibilityAllocations2023, geisslingerEthicalTrajectoryPlanning2023}. 

For simplicity, we consider the same scenario discussed in \cref{sec:BackwardResponsibility} and focus on action selection for just one of the agents. We chose agent 3 as the ego agent, because it has to closely interact with 1 and 2 moving alongside it, with 7 and 8 heading towards it, and with agents 4, 5 and 6 in the cyclical interaction. 

In the backward-looking case, all the actions would have occurred already and one would be looking back at that moment of interaction. However, for the forward-looking case, when choosing its future actions, agent 3  would need to factor in future actions of other agents in planning its own action. In artificial agents, such predictions are typically obtained via a prediction model; here, since prediction models are not the focus of this paper, we assume that agent 3 makes \textit{some} predictions of other agents, and illustrate its reasoning about the resulting FeAR values for a given set of predictions. For simplicity, we assume that these predictions coincide with the actions assigned to these agents in the backward-looking scenario. \cref{fig:BackwardFear_1_scenario}.  

To illustrate how agent 3 could use FeAR-based optimisation over possible actions, we discretised the action space of the ego agent into a finite set of actions parametrised by the magnitude and direction of the acceleration. We then used a grid search to evaluate how different ego actions affect each agent's feasible action space -- something that is quantified by the $\FeAR_{3,j}$ values of agent 3 on other agents (\cref{fig:ForwardFear_1_gridsearch_affected}). Naturally, most actions are courteous towards some agents while being assertive towards others; therefore, to choose one action, the ego agent needs to aggregate the FeAR values across different affected agents. 

To this end, we analysed how the optimal action would change depending on the way of aggregating the FeAR values. One option could be to take the mean of the FeAR values across all agents (\cref{fig:ForwardFear_1_gridsearch_aggregates}, left panel). Using the mean is akin to weighing the advantages and disadvantages to different agents. For reference, \cref{fig:ForwardFear_1_Optimal_action} shows the optimal action obtained by minimising mean FeAR across affected agents while avoiding collisions. Alternatively, if the agent follows Rawlsian fairness and aims to improve the well being of the worst affected, it can choose the optimal action that minimizes the maximum FeAR value caused by an action (\cref{fig:ForwardFear_1_gridsearch_aggregates}, right panel). 
Similarly, if the goal is to maximise the courteousness to any agent, minimum FeAR can be used to find the optimal action (\cref{fig:ForwardFear_1_gridsearch_aggregates}, middle panel).

Instead of the aggregating FeAR \textit{values} across agents, the agent might base its action selection on the \textit{number} of other agents that it's courteous or assertive to (\cref{fig:ForwardFear_1_gridsearch_aggregates_counts}).

Taken together, these results illustrate the richness of the information that the FeAR metric provides. Moreover, they highlight that different choice of the aggregation method can lead to vastly different selected actions.

\section{Discussion}
\label{sec:Discussion}

In this section, we start by discussing the 
assumptions around action-selection and the MdR in spatial interactions~(\cref{sec:Discussion_FeAR_Rationale}).
We then move on to discuss how FeAR compares with other methods for quantifying responsibility~(\cref{sec:Discussion_QuantifyingResponsibility}) and responsible navigation~(\cref{sec:Discussion_ResponsibleNavigation}) to place FeAR within the broader scope of ascribing responsibility and behaving responsibly. 
Finally, we discuss the limitations and potential extensions of the FeAR metric~(\cref{sec:Limitations_FutureWork}). 
For practitioners, we further discuss possible applications~(\cref{sec:Discussion_Applications}).

\subsection{Action selection and the MdR}
\label{sec:Discussion_FeAR_Rationale}

When evaluating whether an action is assertive ($\FeAR_{i,j}>0$)  or courteous ($\FeAR_{i,j}<0$), it is important to consider the normative expectations of how agents are supposed to behave in a given context~\cite{halpernCauseResponsibilityBlame2015, markkulaDefiningInteractionsConceptual2020}. For example, in a supermarket aisle, staying put to look at the shelves is a perfectly acceptable behaviour, while stopping a car in a high-speed lane not. 
The \textit{move de rigueur} (MdR) provides a normative basis for conducting counterfactual analysis of the causal influence of an action. Since, in spatial interactions, all agents are physically embedded in a scene, all of them must perform an action. This means that to study the causal influence of a given action, we need to consider the counterfactual condition of performing another action -- in our case, the MdR. 

MdR represents a normative perspective.
If we consider legality, MdRs can be considered as non-culpable moves which if chosen afford zero responsibility to the actors. From a behavioural lens, MdRs can be considered as a fast and intuitive actions (as described in Kahneman's system 1\cite{kahnemanThinkingFastSlow2011}). From the automation perspective, MdRs can be considered as the default or fail-safe actions that are programmed into automated systems. In an ideal case, MdRs would be designed to align with these perspectives; and also moral, cultural, and other perspectives. Methods like norm synthesis ~\cite{moralesAutomatedSynthesisNormative2013,moralesSynthesisingLiberalNormative2015} or control theoretic approaches \cite{shalev-shwartzFormalModelSafe2018, shalev-shwartzVisionZeroProvable2019, pekVerifyingSafetyLane2017} might be useful for designing MdRs that satisfy traffic rules, societal norms, and constraints posed by human physiological and cognitive limitations~\cite{robolApplyingSocialNorms2016}. 
Nevertheless, designing acceptable MdRs is an open challenge that requires concerted effort from several domains through a participatory approach, similar to designing a moral operational design domain for an automated system \cite{cavalcantesiebertMeaningfulHumanControl2022}.

Using MdR also helps align the FeAR metric with the notion that agents should not be held responsible for events for which they had no avoidance potential \cite{brahamAnatomyMoralResponsibility2012}. In situations where an actor agent has no actions that preclude another agent from colliding ($\nij = 0$ $\forall \action{i}\in \ActionSpace{i}$), the MdR would also lead to the collision ($\nijMdR = 0$), and the FeAR value would be zero.

Although we define MdRs for all agents, they are not obligated to choose these. 

Approaches like Responsible Sensitive Safety (RSS)~\cite{shalev-shwartzFormalModelSafe2018,shalev-shwartzVisionZeroProvable2019} rely on all agents following a recommended action and agents that deviate are considered responsible. 
The FeAR metric offers more agency to individuals by rewarding deviations from the MdR which are beneficial for other agents, instead of penalizing or restraining the choice set. 
This could relax some of the restrictions when defining safe joint-actions in multi-agent systems, while providing a measure of responsibility for all agents.

\subsection{Quantifying Responsibility}
\label{sec:Discussion_QuantifyingResponsibility}

Scholars have followed different methods to model responsibility of agents --- ranging from logical analysis based on the capacity of agents~\cite{loriniLogicalAnalysisResponsibility2014,duijfLogicalStudyMoral2023,yazdanpanahDistantGroupResponsibility2016,yazdanpanahApplyingStrategicReasoning2021}, to probabilistic reasoning about causality~\cite{englTheoryCausalResponsibility2018,chocklerResponsibilityBlameStructuralModel2004,alechina2017causality,triantafyllouActualCausalityResponsibility2022}. 

Epistemic considerations for ascribing blame or praise have also been studied~\cite{loriniLogicalAnalysisResponsibility2014,chocklerResponsibilityBlameStructuralModel2004}.
In this work,  we refrain from epistemic considerations choosing to focus on observable states.
For all forms of responsibility causal contribution is an essential component --- an agent without causal influence cannot be ascribed, praise, blame, moral responsibility or legal liability \cite{vandepoelEthicsTechnologyEngineering2011, vincentStructuredTaxonomyResponsibility2011}. Since causal responsibility is a necessary condition, FeAR can be used to provide a quantitative baseline for determining responsibility which only considers observable behaviours. Once causal responsibility is established, other methods to evaluate epistemic and motivational considerations can be used to ascribe other forms of responsibility.

Another important distinction between FeAR and previous works on quantifying responsibility is that these models pertain to responsibility for an outcome while FeAR pertains to responsibility for the trajectory of the affected agent. In this way, FeAR can provide continuous assessment of responsibility during spatial interactions, not only for a given outcome (e.g., a collision), 
making it particularly tailored for spatial interactions among agents. 

Logical approaches using STIT (Sees-to-it-that) Logic and responsibility games have been used to formalise the responsibility of agents based on whether they have the ability to preclude an outcome~\cite{loriniLogicalAnalysisResponsibility2014,duijfLogicalStudyMoral2023}. These approaches provide a boolean value for whether an agent is responsible or not.  Similar to these, responsibility games under risk have used the conditional probabilities of how different agents act to model responsibility~\cite{duijfLogicalStudyMoral2023}. A logical framework for reasoning about the responsibility of groups of agents have also been proposed which ascribes responsibility for groups based on whether they increase risk beyond a threshold \cite{gladyshevGroupResponsibilityExceeding2023}. All these approaches  provide a boolean value for responsibility while FeAR gives a graded value of causal responsibility.

Another branch of probabilistically reasoning about responsibility incorporates causal models to capture causal relationships between events ~\cite{englTheoryCausalResponsibility2018,bartlingShiftingBlameDelegation2012,chocklerResponsibilityBlameStructuralModel2004,halpernCauseResponsibilityBlame2015,alechina2017causality}.
These models are then used to do counterfactual analysis of how changing an agent's action affects the outcome. Partially Observable Markov Decision Processes (POMDPs) have also been used to model the causal dependencies when reasoning about responsibility~\cite{triantafyllouActualCausalityResponsibility2022}. These methods generally give a graded degree of responsibility as opposed to a boolean value.
These methods are dependent on the causal models and works on the assumption that these models capture the true dynamics of the scenario.
FeAR on the other hand, also provides a graded value of responsibility, while being model agnostic; which means that FeAR is not dependent on assumptions of how agents are likely to behave. 

Approaches with strategic reasoning (e.g., Alternating-time Temporal Logic (ATL) ) about the capacity of groups of agents have been used to model the responsibility of groups of agents for outcomes or failure to deliver outcomes~\cite{loriniLogicalAnalysisResponsibility2014,yazdanpanahDistantGroupResponsibility2016, yazdanpanahApplyingStrategicReasoning2021}. In this work, we focus on single agents instead of groups. However, considering groups of agents is especially important in cases of causal overdetermination as in \cref{fig:CaseStudy12}, where $\FeAR_{1,2}=\FeAR_{3,2}=0$ fails to capture how agents 1 and 3 restrict agent 2. Even though we do not consider all possible groups of agents, the diagonal values ($\FeAR_{j,j}$) are able to provide insights into how the group of all other agents $\neg j$ affects agent $j$; as seen in \cref{fig:CaseStudy12}, where $\FeAR_{2,2}=0$ shows how the other agents are completely restricting agent 2.

In the context of human-AI collaboration, an information theoretic model of comparative human causal responsibility has been proposed based on how much of the outcome is dependent on the variability introduced by the human~
\cite{douerJudgingOneOwn2022, douerResponsibilityQuantificationModel2020, douerTheoreticalMeasuredSubjective2021}. This model provides a dichotomy of responsibility for the human and the automation. 
FeAR, on the other hand, treats all agents equally, whether they are automated or human, and can thus be applied to reason about causal responsibility in human-automation interactions. Quantifying responsibility for humans and autonomous agents is an ongoing challenge in the meaningful human control community \cite{mecacciResearchHandbookMeaningful2024}. Such quantification is crucial for operationalizing the philosophical concept of meaningful human control \cite{santonidesioMeaningfulHumanControl2018}, beyond listing necessary properties for evaluating, or designing systems under meaningful human control \cite{cavalcantesiebertMeaningfulHumanControl2022}. For example, FeAR - being agnostic about whether an agent is human or not - can be readily applied to mixed-traffic interactions with human, automated and hybrid vehicles --- which makes it a great tool for envisioning a new ethics of transportation~\cite{santonidesioEuropeanCommissionReport2021, papadimitriouCommonEthicalSafe2022} and ensuring meaningful human control over automated mobility~\cite{calvertDesigningAutomatedVehicle2023}.  

The FeAR metric was originally proposed in a grid-world setting with discrete actions and states~\cite{georgeFeasibleActionSpaceReduction2023a}. Similarly, all the above models of responsibility pertain to discrete actions. Therefore applying these models to spatial interactions with continuous actions is problematic as the computational complexity increases as the number of actions to consider increase. In this paper, we partition the action space into subsets to manage the computational complexity of quantifying the feasible action space. Thus, the current formulation of the FeAR metric can be applied to continuous action spaces encountered in spatial interactions.

Hence, compared to the fundamental and computational models of responsibility, the FeAR metric is able to provide a model-agnostic and graded measure of causal responsibility for the trajectory of spatially interacting agents with continuous actions.

\subsection{Responsible Navigation}
\label{sec:Discussion_ResponsibleNavigation}

Responsible navigation has also garnered much interest in relation to autonomous vehicles; including control-theoretic~\cite{shalev-shwartzFormalModelSafe2018, shalev-shwartzVisionZeroProvable2019}, data-driven~\cite{cosnerLearningResponsibilityAllocations2023,remyLearningResponsibilityAllocations2024} and planning-based~\cite{geisslingerEthicalTrajectoryPlanning2023,pekVerifyingSafetyLane2017} approaches to navigate responsibly around other road users.  

Responsible Sensitive Safety (RSS) is a control-theoretic approach that formalises the ``Duty of Care'' and prescribes ``proper responses'' for vehicles, which if followed by all vehicles makes accidents rare \cite{shalev-shwartzFormalModelSafe2018, shalev-shwartzVisionZeroProvable2019}. RSS has meticulously modelled various traffic scenarios and prescribed actions that are verifiably safe. For example, RSS limits the maximum deceleration of vehicles so that vehicles that follow it at a minimum safe distance have sufficient time and space to execute evasive manoeuvres. According to their model, if a collision happens, a vehicle that violated the ``proper response'' of RSS will be held responsible. This however can lead to problematic ascriptions of responsibility when a vehicle deviates from the proper response of RSS but ends up being being beneficial to one or more road users. If we take the proper response of the RSS as the Move de Rigueur, we can use FeAR to analyse how assertive or courteous the actions of an agent are --- as opposed to considering all deviations as negative.

Another data-driven approach to responsible navigation involves learning Responsibility-Aware Control Barrier Functions (RA-CBFs) to model which agent should deviate more from their optimal path in a given context \cite{cosnerLearningResponsibilityAllocations2023,remyLearningResponsibilityAllocations2024}. According to this approach, agents that deviate more from their optimal trajectory and make room for others are considered to be acting responsibly. This model does not explicitly consider the philosophical notion of responsibility and as such cannot be used to ascribe responsibility to past interactions. FeAR models causal responsibility in spatial interactions and can thus be included in the training of RA-CBFs to better inform the model about the actual causal responsibility of agents. The current formulation of FeAR which considers all possible actions of all agents in the scene is computationally expensive and is not ideal for real-time applications. Learning-based approaches like RA-CBFs which take FeAR as ground truth while training, could be used to generate real-time responsibility-aware navigation policies for agents.

Yet another approach to ethical trajectory planning proposed using reachable sets of road users to quantify the risk of possible collisions which in turn was used to model responsibility for such collisions~\cite{geisslingerEthicalTrajectoryPlanning2023,pekVerifyingSafetyLane2017}. The reachable sets describe where an agent could be at a particular time given certain physical constraints and traffic rules. If an agent $j$ collides with another agent $i$ within the reachable set of $j$, then $i$ would be ascribed more responsibility. Probabilistic models of the behaviour of agents are necessary to calculate the risk associated with possible collisions. FeAR on the other hand does not rely on such probabilities and provides a more objective measure for how one agent affects another. Furthermore, when we consider the reachable set, we are looking only at the terminal configurations at a given time and do not explicitly consider the different trajectories that could have led there. FeAR on the other hand considers the feasible actions of the agents and thereby accounts for the feasibility of the entire trajectory and not just the terminal location. It is true that in our current parametrisation of the action space based on constant acceleration, each terminal state corresponds to only one action. But, a different parametrisation of the action space (like a linear interpolation between two accelerations at the start and end of the time window), could lead to multiple paths that lead to the same terminal state. Thus, the feasible action space better represents the agency available to agents than the reachable set. 

Ergo, compared to the more applied methods for responsible navigation, FeAR offers a more rigorous account of causal responsibility and incorporating FeAR into the training of such applied methods could lead to fundamentally rooted methods for responsible navigation.

\subsection{Limitations and Future Work}
\label{sec:Limitations_FutureWork}

To ascribe praise, blame, liability or moral responsibility, in addition to causal influence, considerations must be made for the epistemic and motivational characteristics of actors~ \cite{vandepoelEthicsTechnologyEngineering2011}. In this work, we abstain from epistemic and motivational considerations and focus observable behaviours to quantify causal responsibility.
Causal responsibility is just one necessary condition and further deliberation is needed to hold someone, blameworthy, praiseworthy, liable or morally responsible.
For example, given a large dataset of traffic interactions, we can use the FeAR metric to flag scenarios with high values of FeAR. The flagged scenarios could then be evaluated by human experts or automated agents to check for epistemic, motivational and legal concerns.
Designing frameworks that combine FeAR with deliberations on empirical, motivational and legal considerations to evaluate responsibility are fertile grounds for future research.

Like all other models that ignore group responsibility, FeAR fails in cases of causal overdetermination where more than one agent completely restrict the feasible actions of an affected agent (as illustrated in \cref{fig:CaseStudy12}). To address causal overdetermination we need models of collective causal responsibility\cite{yazdanpanahDistantGroupResponsibility2016, yazdanpanahDifferentFormsResponsibility2021, yazdanpanahApplyingStrategicReasoning2021}. 
Future research could explore how the FeAR metric can be adapted to account for causal influence of groups of agents on an affected agent.

Currently, when quantifying the causal responsibility during a time window given some initial state, we ignore the history of states and actions that led to this initial state. To get a complete picture of how past actions of agents affect the trajectories of agents, future work needs formulations of FeAR applicable to sequences of actions.

In this work. we presented FeAR using point-mass dynamics and ignored the orientation of agents (which is applicable to pedestrians and robots with holonomic constraints) to highlight the utility of the FeAR metric, without simultaneously addressing the complexity of non-holonomic trajectories
To apply FeAR to vehicles and non-holonomic robots, the future work must account for the orientation of agents and their non-holonomic constraints.

In summary, future work on FeAR could address epistemic and motivational considerations, collective responsibility, responsibility over sequence of actions and non-holonomic dynamics.

\subsection{The use of FeAR}
\label{sec:Discussion_Applications}

For practitioners in the domain of spatial interactions, we reflect on the potential applications of the FeAR metric for backward-looking and forward-looking responsibility, and as a surrogate measure of safety.

\noindent\textbf{Backward-looking responsibility:}
If we have the trajectories of agents, we can find the actions taken by agents at each time instant and the FeAR values calculated using these actions will quantify the backward-looking causal responsibility for the trajectory of agents for a specified time window.  
Considerations like the limits of the abilities of agents and the planning time horizon of agents should be factored into the choice of appropriate parameters for getting meaningful estimates of causal responsibility.  

\noindent\textbf{Forward-Looking Responsibility:}
For forward-looking responsibility, since we do not know the future actions of agents, prediction models~\cite{rudenkoHumanMotionTrajectory2020} should be used to get predicted actions of agents, which can then be used to compute FeAR values. Given some predicted joint action of other agents, FeAR values can be used to score different ego actions. Thus, to improve the courteousness of agents, FeAR can be incorporated into the cost functions of ethical trajectory planning algorithms; like the one that evenly distributes risk amongst
different road users~\cite{geisslingerEthicalTrajectoryPlanning2023}.

\noindent\textbf{Surrogate Measures of Safety:}
Many surrogate measures of safety have been widely used by traffic scientists to study traffic interactions \cite{chenEvaluatingSafetyEfficiency2024,mullakkal-babuProbabilisticFieldApproach2020}. Most of these measures like time to arrival, time to collision, time to lane change, post encroachment time, lateral distance, headway, etc. are specified only for particular contexts. For example, time to collision and post encroachment time are not defined for agents moving parallel to each other. This lack of universality of metrics can complicate how studies about traffic interactions can be generalised to different contexts.
A data-driven approach to quantify safety across different context was based on how the distance between vehicles compares to the expected distance in cases with similar velocities~\cite{jiaoUnifiedProbabilisticApproach2025}. The FeAR metric adds another dimension to this discussion and can be used as a surrogate metric for safety, derived from first principles, which can be applied across different contexts.

\section{Conclusion}
\label{sec:Conclusion}
We formulated the feasible action-space reduction (FeAR) metric to quantify causal responsibility in spatial interactions with actions in the continuous domain. Through case studies with prototypical space-sharing conflicts, we illustrated the properties of the metric and how it is sensitive to actions of agents, group interactions, and normative expectations represented via the \textit{Moves de Rigueur} (MdR). We applied the metric to calculate backward-looking and forward-looking responsibility, and showed one potential approach to incorporate FeAR for responsibility-aware motion planning. 
FeAR is able to capture how individual agents can be assertive to some agents, i.e., more causal responsibility for the trajectory of the affected and courteous to others i.e., less causal responsibility. The magnitude of the FeAR values also helps to compare the causal responsibility of 1) different agents on an affected agent or, 2) one agent on different affected agents. Thus, the FeAR metric along with the MdR capture the causal responsibility of one agent's action on the trajectory of another agent for a time window of spatial interaction with concurrent action selection.


\section*{Declarations}

\noindent\textbf{Author Contributions} Ashwin George conducted the research and was supervised by Arkady Zgonnikov, Luciano Cavalcante Siebert and David A. Abbink. All authors contributed to the first draft of the paper and subsequent revisions.

\noindent\textbf{Funding}
This work is supported by the TU Delft AI Labs programme.

\noindent\textbf{Acknowledgements}
The authors would like to thank Chris Pek for valuable suggestions for the implementation of collision checks for spatial interactions. 

\noindent\textbf{Competing interests}
The authors have no conflicts of interest to declare that are relevant to the content of this article.

\noindent\textbf{Data availability}
The code for modelling and analysing case studies in the study is available at \url{https://github.com/DAI-Lab-HERALD/continuousFeAR}.
The data generated during the analysis of case studies can be found at: \url{https://osf.io/8mtnf/}.

\appendix
\onecolumn

\newpage
\section{Collision checks}
\begin{figure}[htb]
    \centering
    \begin{subfigure}{0.44\linewidth}
        \includegraphics[width=\linewidth]{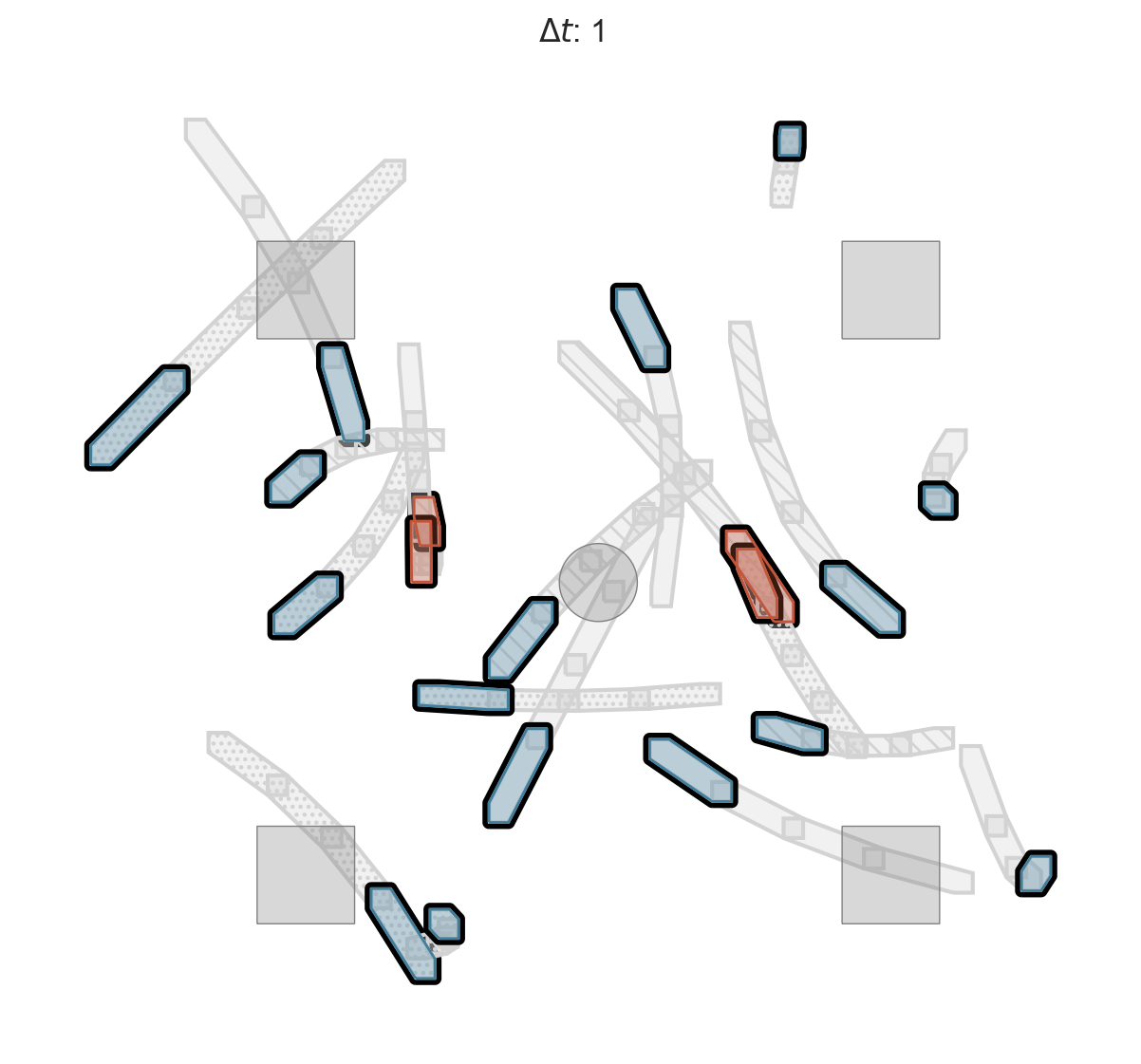}
    \end{subfigure}
    \begin{subfigure}{0.44\linewidth}
        \includegraphics[width=\linewidth]{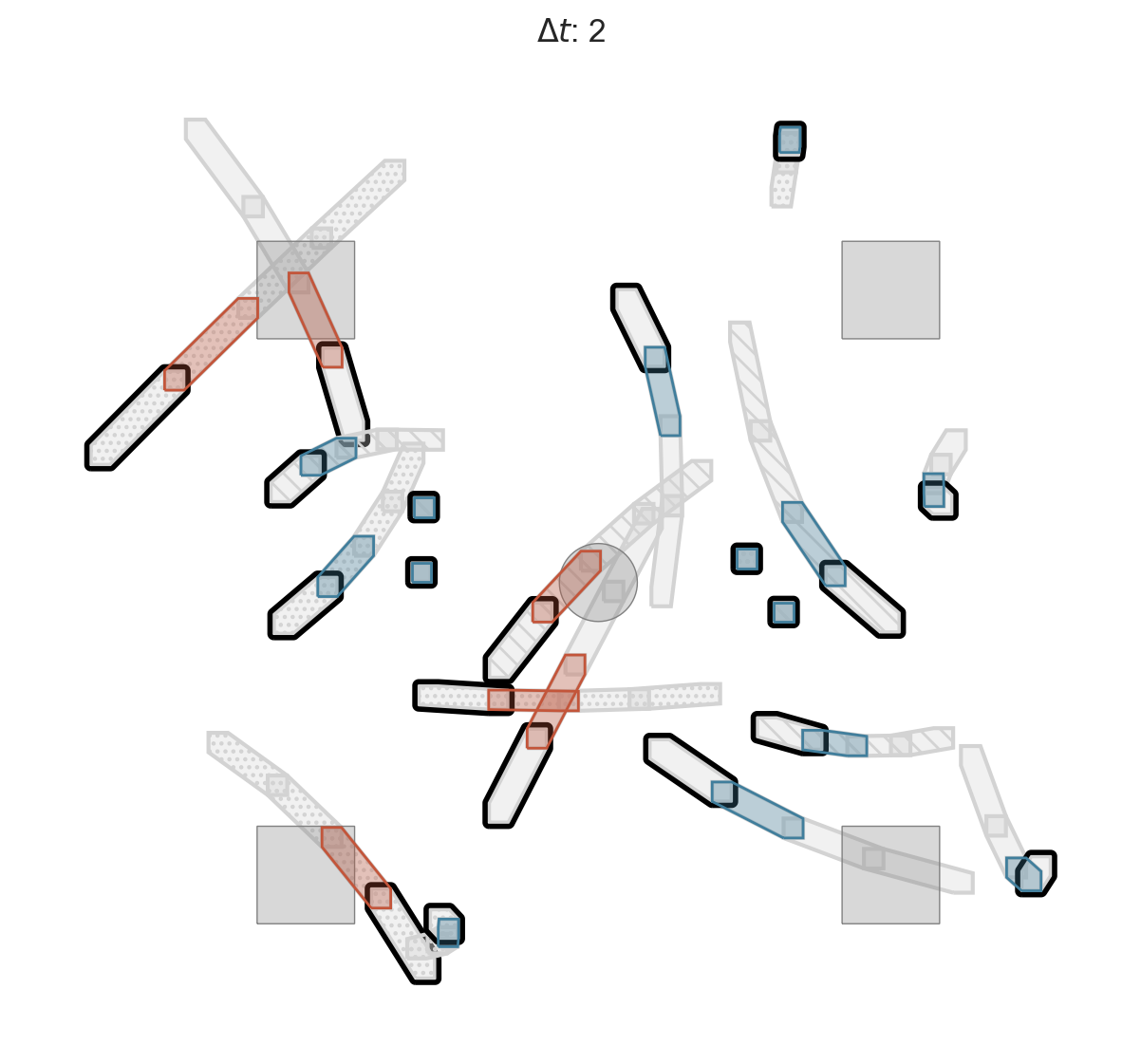}
    \end{subfigure}
    \begin{subfigure}{0.44\linewidth}
        \includegraphics[width=\linewidth]{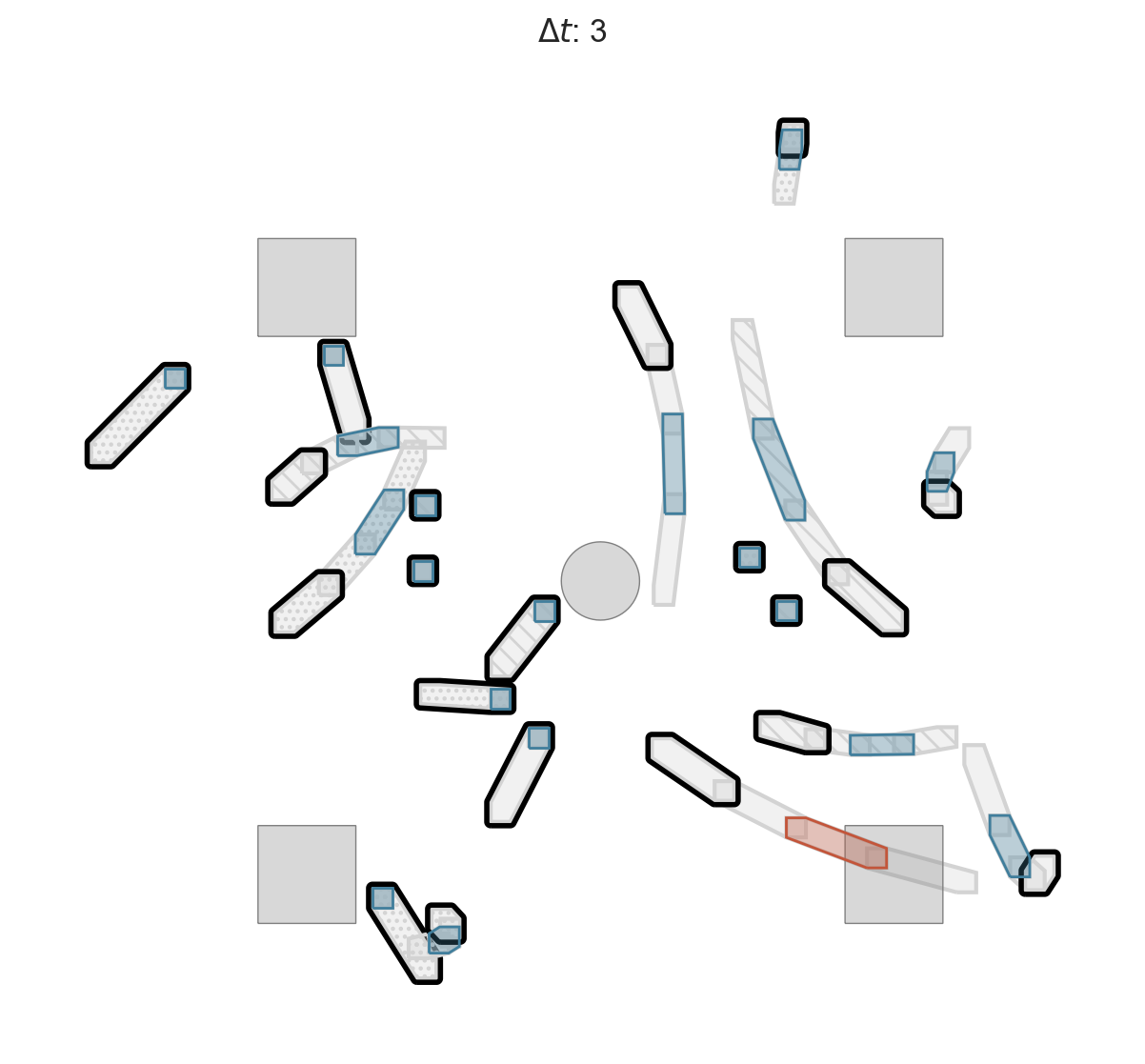}
    \end{subfigure}
    \begin{subfigure}{0.44\linewidth}
        \includegraphics[width=\linewidth]{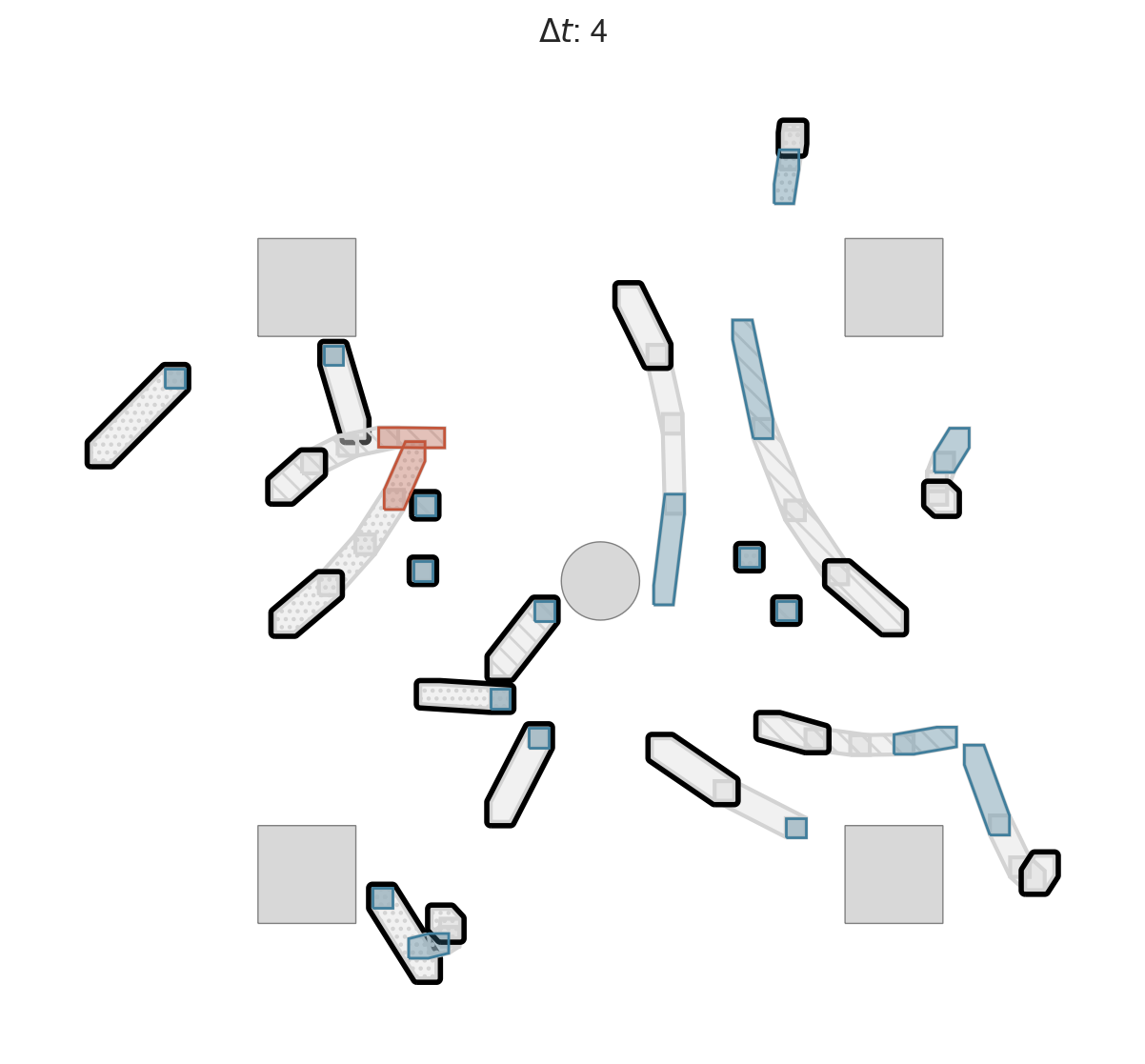}
    \end{subfigure}
    \caption{Here we show four consecutive time intervals where colliding agents are shown in red. Each colliding agent in considered stationary in subsequent time intervals.}
    \label{fig:traj_hulls_collsion_checks}
\end{figure}

As discussed in \cref{sec:Preliminaries}, square bounding boxes around agents were used to check for collisions. The time window in consideration was divided into time intervals and the the trajectory hulls formed by the starting and final locations of agents during that time interval were used to check for collisions with other agents or static obstacles as shown in \cref{fig:traj_hulls_collsion_checks}. Collision checks were done starting from the first time interval and colliding agents were considered stationary in subsequent time intervals, reverting to their last collision free location.

\newpage

\section{\textit{Moves de Rigueur} based on Social Force Models}

To illustrate the application of \textit{move de rigueur} (MdR), we presented a candidate policy for computing the MdR based on the social force model for pedestrian interactions~\cite{helbingSocialForceModel1995}. According to this model, humans are repelled away from each other by a force that is inversely proportional to the distance between them while being attracted to some goal. We modelled this social force as per \cref{alg:socialForceAcceleration} and used this to compute MdRs for agents as in \cref{alg:getMDRTarget}. For more details on the implementation, please refer to the Github repository.

\begin{algorithm}
    \caption{Compute Social Force Acceleration}
    \label{alg:socialForceAcceleration}
    \begin{algorithmic}[1]
        \Procedure{socialForceAcceleration}{x, y, k, \texttt{threshold\_distance}, \texttt{buffer}, \texttt{threshold\_a}}
        \State \textbf{Input:} $x, y$ (positions of agents), $k$ (force constant), \texttt{threshold\_distance}, \texttt{buffer}, \texttt{threshold\_a}.
        \State \textbf{Initialize:} $\texttt{distances} \gets \sqrt{(x_i - x_j)^2 + (y_i - y_j)^2}$ for all $i, j$.
        
        \State \textbf{Ensure:} $\texttt{distances}[i, i] \gets 1e-5$ \Comment{Avoid self-distance becoming zero.}
        \State \textbf{Ensure:} $\texttt{distances}[i, j] \geq \texttt{buffer}$ \Comment{Enforce buffer for collision avoidance.}
        
        \State $\text{inv\_sq\_distances} \gets \frac{1}{(\texttt{distances} - \texttt{buffer})^2}$
        \State $\text{inv\_sq\_distances}[\texttt{distances} > \texttt{threshold\_distance}] \gets 0$ \Comment{Mask forces beyond the threshold.}

        \State $\Delta x \gets x_i - x_j$ \Comment{X-distance between agents.}
        \State $\Delta y \gets y_i - y_j$ \Comment{Y-distance between agents.}

        \State $\hat{r}_x \gets \frac{\Delta x}{\texttt{distances}}$, $\hat{r}_y \gets \frac{\Delta y}{\texttt{distances}}$ \Comment{Unit vectors for direction.}

        \State $a_\text{social\_x} \gets k \cdot \hat{r}_x \cdot \text{inv\_sq\_distances}$
        \State $a_\text{social\_y} \gets k \cdot \hat{r}_y \cdot \text{inv\_sq\_distances}$
        
        \State $(a, \theta) \gets \textsc{getAandTheta}(a_\text{social\_x}, a_\text{social\_y})$ \Comment{Compute acceleration magnitude and direction.}

        \State $a \gets \text{clip}(a, 0, \texttt{threshold\_a})$ \Comment{Clip acceleration to the max threshold.}
        
        \State $a_\text{social\_x} \gets a \cdot \cos(\theta)$
        \State $a_\text{social\_y} \gets a \cdot \sin(\theta)$

        \State \textbf{Ensure:} $\text{Set diagonal elements of } a_\text{social\_x}, a_\text{social\_y} \gets 0$ \Comment{No self-interaction.}

        \State \textbf{Output:} $\sum a_\text{social\_x}, \sum a_\text{social\_y}, \texttt{distances}$
        \EndProcedure
    \end{algorithmic}
\end{algorithm}
\begin{algorithm}
    \caption{Compute Moves de Rigueurs (MDR) for Agents}
    \label{alg:getMDRTarget}
    \begin{algorithmic}[1]
        \Procedure{get-MDR-Target}{}
        \State \textbf{Input:}\parbox[t]{0.45\linewidth}{\raggedright $x, y, v0x, v0y, \texttt{time\_per\_step}$, \texttt{vx\_desired}, \texttt{vy\_desired}, $k_v$, $k$, \texttt{threshold\_distance}, \texttt{threshold\_velocity}, \texttt{restore\_factor}, \texttt{lanes}, $k_\text{lane}$.}
        \State \textbf{Initialize:} $x, y, v0x, v0y, \texttt{vx\_desired}, \texttt{vy\_desired}$.
        
        \State $(a_\text{social\_x}, a_\text{social\_y}, \texttt{distances}) \gets \parbox[t]{0.45\linewidth}{\textsc{socialForceAcceleration}(x, y, k, \texttt{threshold\_distance}, \texttt{buffer})}$
        \State $\text{Set diagonal elements of } \texttt{distances} \gets \infty$
        \State $\texttt{min\_distances} \gets \textsc{min}(\texttt{distances}, \text{axis}=1)$

        \For{\textbf{each agent}}
            \State $v_x \gets a_\text{social\_x} \cdot \texttt{time\_per\_step}$
            \State $v_y \gets a_\text{social\_y} \cdot \texttt{time\_per\_step}$
            \State $\Delta v_x \gets v0x - v_x$
            \State $\Delta v_y \gets v0y - v_y$
            \State $\Delta_\text{desired\_vx} \gets \texttt{vx\_desired} - v0x$
            \State $\Delta_\text{desired\_vy} \gets \texttt{vy\_desired} - v0y$
            \State $a_x \gets a_\text{social\_x} + \Delta v_x \cdot \texttt{restore\_factor} + k_v \cdot \Delta_\text{desired\_vx}$
            \State $a_y \gets a_\text{social\_y} + \Delta v_y \cdot \texttt{restore\_factor} + k_v \cdot \Delta_\text{desired\_vy}$
            \State $v_x \gets v0x + a_x \cdot \texttt{time\_per\_step}$
            \State $v_y \gets v0y + a_y \cdot \texttt{time\_per\_step}$
        \EndFor

        \State $v_\text{max} \gets \min\left(\frac{\texttt{min\_distances} / 2 - \texttt{buffer} / 2}{\texttt{time\_per\_step}}, \texttt{threshold\_velocity}\right)$
        \State $v \gets \text{clip}(v, 0, v_\text{max})$
        \State $v_x \gets v \cdot \cos(\theta_v)$
        \State $v_y \gets v \cdot \sin(\theta_v)$

        \State $a_x \gets \frac{v_x - v0x}{\texttt{time\_per\_step}}$
        \State $a_y \gets \frac{v_y - v0y}{\texttt{time\_per\_step}}$

        \For{\textbf{each} \texttt{lane} in \texttt{lanes}}
            \For{\textbf{each agent}}
                \If{\texttt{lane polynomial exists}}
                    \State $(a_\text{lane}, \theta_\text{lane}) \gets \textsc{accelerateToLane}(x, y, v0x, v0y, \texttt{lane})$
                    \State $a_x \gets \frac{a_x + k_\text{lane} \cdot a_\text{lane} \cdot \cos(\theta_\text{lane})}{1 + k_\text{lane}}$
                    \State $a_y \gets \frac{a_y + k_\text{lane} \cdot a_\text{lane} \cdot \sin(\theta_\text{lane})}{1 + k_\text{lane}}$
                \EndIf
            \EndFor
        \EndFor

        \State \textbf{Output:} $a, \theta$
        \EndProcedure
    \end{algorithmic}
\end{algorithm}

\newpage

\bibliography{cFeAR}

\begin{thebibliography}{10}

\bibitem{dignumResponsibleArtificialIntelligence2019}
V.~Dignum, {\em Responsible {{Artificial Intelligence}}: {{How}} to {{Develop}} and {{Use AI}} in a {{Responsible Way}}}.
\newblock Artificial {{Intelligence}}: {{Foundations}}, {{Theory}}, and {{Algorithms}}, Cham: Springer International Publishing, 2019.

\bibitem{dastaniResponsibilityAISystems2023}
M.~Dastani and V.~Yazdanpanah, ``Responsibility of {{AI Systems}},'' {\em AI \& SOCIETY}, vol.~38, pp.~843--852, Apr. 2023.

\bibitem{europeancommission.directorategeneralforcommunicationsnetworkscontentandtechnology.EthicsGuidelinesTrustworthy2019}
{European Commission. Directorate General for Communications Networks, Content and Technology.} and {High Level Expert Group on Artificial Intelligence.}, {\em Ethics Guidelines for Trustworthy {{AI}}.}
\newblock LU: Publications Office, 2019.

\bibitem{beckersDriversPartiallyAutomated2022}
N.~Beckers, L.~C. Siebert, M.~Bruijnes, C.~Jonker, and D.~Abbink, ``Drivers of partially automated vehicles are blamed for crashes that they cannot reasonably avoid,'' {\em Scientific Reports}, vol.~12, p.~16193, Sept. 2022.

\bibitem{papadimitriouCommonEthicalSafe2022}
E.~Papadimitriou, H.~Farah, G.~{van de Kaa}, F.~{Santoni de Sio}, M.~Hagenzieker, and P.~{van Gelder}, ``Towards common ethical and safe `behaviour' standards for automated vehicles,'' {\em Accident Analysis \& Prevention}, vol.~174, p.~106724, Sept. 2022.

\bibitem{santonidesioEuropeanCommissionReport2021}
F.~{Santoni de Sio}, ``The {{European Commission}} report on ethics of connected and automated vehicles and the future of ethics of transportation,'' {\em Ethics and Information Technology}, vol.~23, pp.~713--726, Dec. 2021.

\bibitem{calvertDesigningAutomatedVehicle2023}
S.~C. Calvert, S.~O. Johnsen, and A.~George, ``Designing automated vehicle and traffic systems towards meaningful human control,'' in {\em Research Handbook on Meaningful Human Control of Artificial Intelligence Systems}, United Kingdom: Edward Elgar Publishing, 2023 (Accepted).

\bibitem{dignum2020agents}
V.~Dignum and F.~Dignum, ``Agents are dead. {{Long}} live agents!,'' in {\em Proceedings of the 19th International Conference on Autonomous Agents and {{MultiAgent}} Systems}, pp.~1701--1705, 2020.

\bibitem{yazdanpanahReasoningResponsibilityAutonomous2022}
V.~Yazdanpanah, E.~H. Gerding, S.~Stein, M.~Dastani, C.~M. Jonker, T.~J. Norman, and S.~D. Ramchurn, ``Reasoning about responsibility in autonomous systems: Challenges and opportunities,'' {\em AI \& SOCIETY}, Dec. 2022.

\bibitem{vincentStructuredTaxonomyResponsibility2011}
N.~A. Vincent, ``A {{Structured}} taxonomy of responsibility concepts,'' in {\em Moral Responsibility} (N.~A. Vincent, I.~{van de Poel}, and J.~{van den Hoven}, eds.), Library of Ethics and Applied Philosophy, pp.~15--35, United States: Springer, Springer Nature, 2011.

\bibitem{hartCausationLaw2002}
H.~L.~A. Hart and T.~Honor{\'e}, {\em Causation in the Law}.
\newblock Oxford: Clarendon Press, 2. ed., repr~ed., 2002.

\bibitem{vandepoelEthicsTechnologyEngineering2011}
I.~R. {van de Poel} and L.~M. Royakkers, {\em Ethics, Technology, and Engineering : An Introduction}.
\newblock United States: Wiley-Blackwell, 2011.

\bibitem{shalev-shwartzFormalModelSafe2018}
S.~{Shalev-Shwartz}, S.~Shammah, and A.~Shashua, ``On a {{Formal Model}} of {{Safe}} and {{Scalable Self-driving Cars}},'' Oct. 2018.

\bibitem{cosnerLearningResponsibilityAllocations2023}
R.~K. Cosner, Y.~Chen, K.~Leung, and M.~Pavone, ``Learning responsibility allocations for safe human-robot interaction with applications to autonomous driving,'' in {\em Proc. {{IEEE}} Conf. on Robotics and Automation}, 2023.

\bibitem{geisslingerEthicalTrajectoryPlanning2023}
M.~Geisslinger, F.~Poszler, and M.~Lienkamp, ``An ethical trajectory planning algorithm for autonomous vehicles,'' {\em Nature Machine Intelligence}, vol.~5, pp.~137--144, Feb. 2023.

\bibitem{remyLearningResponsibilityAllocations2024}
I.~Remy, D.~{Fridovich-Keil}, and K.~Leung, ``Learning responsibility allocations for multi-agent interactions: {{A}} differentiable optimization approach with control barrier functions,'' Oct. 2024.

\bibitem{shalev-shwartzVisionZeroProvable2019}
S.~{Shalev-Shwartz}, S.~Shammah, and A.~Shashua, ``Vision {{Zero}}: On a {{Provable Method}} for {{Eliminating Roadway Accidents}} without {{Compromising Traffic Throughput}},'' Jan. 2019.

\bibitem{loriniLogicalAnalysisResponsibility2014}
E.~Lorini, D.~Longin, and E.~Mayor, ``A logical analysis of responsibility attribution: Emotions, individuals and collectives,'' {\em Journal of Logic and Computation}, vol.~24, pp.~1313--1339, Dec. 2014.

\bibitem{duijfLogicalStudyMoral2023}
H.~Duijf, ``A {{Logical Study}} of {{Moral Responsibility}},'' {\em Erkenntnis}, Sept. 2023.

\bibitem{gladyshevGroupResponsibilityExceeding2023}
M.~Gladyshev, N.~Alechina, M.~Dastani, and D.~Doder, ``Group {{Responsibility}} for {{Exceeding Risk Threshold}},'' in {\em Proceedings of the {{Twentieth International Conference}} on {{Principles}} of {{Knowledge Representation}} and {{Reasoning}}}, (Rhodes, Greece), pp.~322--332, International Joint Conferences on Artificial Intelligence Organization, Sept. 2023.

\bibitem{yazdanpanahApplyingStrategicReasoning2021}
V.~Yazdanpanah, S.~Stein, E.~H. Gerding, and N.~R. Jennings, ``Applying strategic reasoning for accountability ascription in multiagent teams,'' in {\em Proceedings of the Workshop on Artificial Intelligence Safety 2021 Co-Located with the Thirtieth International Joint Conference on Artificial Intelligence ({{IJCAI}} 2021), Virtual, August, 2021} (H.~Espinoza, J.~A. McDermid, X.~Huang, M.~{Castillo-Effen}, X.~C. Chen, J.~{Hern{\'a}ndez-Orallo}, S.~{\'O}. {h{\'E}igeartaigh}, R.~Mallah, and G.~Pedroza, eds.), vol.~2916 of {\em {{CEUR}} Workshop Proceedings}, CEUR-WS.org, 2021.

\bibitem{yazdanpanahDistantGroupResponsibility2016}
V.~Yazdanpanah and M.~Dastani, ``Distant {{Group Responsibility}} in {{Multi-agent Systems}},'' in {\em {{PRIMA}} 2016: {{Princiles}} and {{Practice}} of {{Multi-Agent Systems}}} (M.~Baldoni, A.~K. Chopra, T.~C. Son, K.~Hirayama, and P.~Torroni, eds.), vol.~9862, pp.~261--278, Cham: Springer International Publishing, 2016.

\bibitem{englTheoryCausalResponsibility2018}
F.~Engl, ``A {{Theory}} of {{Causal Responsibility Attribution}},'' {\em SSRN Electronic Journal}, 2018.

\bibitem{halpernCausesExplanationsStructuralModel2005}
J.~Y. Halpern and J.~Pearl, ``Causes and {{Explanations}}: {{A Structural-Model Approach}}. {{Part I}}: {{Causes}},'' {\em The British Journal for the Philosophy of Science}, vol.~56, pp.~843--887, Dec. 2005.

\bibitem{bartlingShiftingBlameDelegation2012}
B.~Bartling and U.~Fischbacher, ``Shifting the blame: {{On}} delegation and responsibility,'' {\em The Review of Economic Studies}, vol.~79, no.~1, pp.~67--87, 2012.

\bibitem{chocklerResponsibilityBlameStructuralModel2004}
H.~Chockler and J.~Y. Halpern, ``Responsibility and {{Blame}}: {{A Structural-Model Approach}},'' {\em Journal of Artificial Intelligence Research}, vol.~22, pp.~93--115, Oct. 2004.

\bibitem{halpernCauseResponsibilityBlame2015}
J.~Y. Halpern, ``Cause, responsibility and blame: A structural-model approach,'' {\em Law, Probability and Risk}, vol.~14, pp.~91--118, June 2015.

\bibitem{triantafyllouActualCausalityResponsibility2022}
S.~Triantafyllou, A.~Singla, and G.~Radanovic, ``Actual {{Causality}} and {{Responsibility Attribution}} in {{Decentralized Partially Observable Markov Decision Processes}},'' in {\em Proceedings of the 2022 {{AAAI}}/{{ACM Conference}} on {{AI}}, {{Ethics}}, and {{Society}}}, (Oxford United Kingdom), pp.~739--752, ACM, July 2022.

\bibitem{alechina2017causality}
N.~Alechina, J.~Y. Halpern, and B.~Logan, ``Causality, {{Responsibility}} and {{Blame}} in {{Team Plans}},'' in {\em Proceedings of the 16th Conference on Autonomous Agents and {{MultiAgent}} Systems}, pp.~1091--1099, 2017.

\bibitem{douerResponsibilityQuantificationModel2020}
N.~Douer and J.~Meyer, ``The {{Responsibility Quantification Model}} of {{Human Interaction With Automation}},'' {\em IEEE Transactions on Automation Science and Engineering}, vol.~17, pp.~1044--1060, Apr. 2020.

\bibitem{douerJudgingOneOwn2022}
N.~Douer and J.~Meyer, ``Judging {{One}}'s {{Own}} or {{Another Person}}'s {{Responsibility}} in {{Interactions With Automation}},'' {\em Human Factors: The Journal of the Human Factors and Ergonomics Society}, vol.~64, pp.~359--371, Mar. 2022.

\bibitem{georgeFeasibleActionSpaceReduction2023a}
A.~George, L.~Cavalcante~Siebert, D.~Abbink, and A.~Zgonnikov, ``Feasible {{Action-Space Reduction}} as a {{Metric}} of {{Causal Responsibility}} in {{Multi-Agent Spatial Interactions}},'' in {\em Frontiers in {{Artificial Intelligence}} and {{Applications}}} (K.~Gal, A.~Now{\'e}, G.~J. Nalepa, R.~Fairstein, and R.~R{\u a}dulescu, eds.), vol.~372 of {\em Frontiers in {{Artificial Intelligence}} and {{Applications}}}, pp.~819--826, IOS Press, Sept. 2023.

\bibitem{durraniNewCarfollowingModel2024}
U.~Durrani and C.~Lee, ``A new car-following model with incorporation of {{Markkula}}'s framework of sensorimotor control in sustained motion tasks,'' {\em Transportation Research Part B: Methodological}, vol.~184, p.~102969, June 2024.

\bibitem{gawthropIntermittentControlComputational2011}
P.~Gawthrop, I.~Loram, M.~Lakie, and H.~Gollee, ``Intermittent control: A computational theory of human control,'' {\em Biological Cybernetics}, vol.~104, pp.~31--51, Feb. 2011.

\bibitem{markkulaSustainedSensorimotorControl2018}
G.~Markkula, E.~Boer, R.~Romano, and N.~Merat, ``Sustained sensorimotor control as intermittent decisions about prediction errors: Computational framework and application to ground vehicle steering,'' {\em Biological Cybernetics}, vol.~112, pp.~181--207, June 2018.

\bibitem{siebenCollectivePhenomenaCrowds2017}
A.~Sieben, J.~Schumann, and A.~Seyfried, ``Collective phenomena in crowds---{{Where}} pedestrian dynamics need social psychology,'' {\em PLOS ONE}, vol.~12, p.~e0177328, June 2017.

\bibitem{parnellResilientInteractionsCyclists2024}
K.~J. Parnell, S.~E. Merriman, and K.~L. Plant, ``Resilient interactions between cyclists and drivers, and what does this mean for automated vehicles?,'' {\em Applied Ergonomics}, vol.~117, p.~104237, May 2024.

\bibitem{bjorklundDriverBehaviourIntersections2005}
G.~M. Bj{\"o}rklund and L.~{\AA}berg, ``Driver behaviour in intersections: {{Formal}} and informal traffic rules,'' {\em Transportation Research Part F: Traffic Psychology and Behaviour}, vol.~8, pp.~239--253, May 2005.

\bibitem{robolApplyingSocialNorms2016}
M.~Robol, P.~Giorgini, and P.~Busetta, ``Applying social norms to high-fidelity pedestrian and traffic simulations,'' in {\em 2016 {{IEEE International Smart Cities Conference}} ({{ISC2}})}, (Trento, Italy), pp.~1--6, IEEE, Sept. 2016.

\bibitem{chenSocialLearningMarkov2022}
X.~Chen, Z.~Li, and X.~Di, ``Social {{Learning In Markov Games}}: {{Empowering Autonomous Driving}},'' in {\em 2022 {{IEEE Intelligent Vehicles Symposium}} ({{IV}})}, (Aachen, Germany), pp.~478--483, IEEE, June 2022.

\bibitem{helbingSocialForceModel1995}
D.~Helbing and P.~Moln{\'a}r, ``Social force model for pedestrian dynamics,'' {\em Physical Review E}, vol.~51, pp.~4282--4286, May 1995.

\bibitem{markkulaDefiningInteractionsConceptual2020}
G.~Markkula, R.~Madigan, D.~Nathanael, E.~Portouli, Y.~M. Lee, A.~Dietrich, J.~Billington, A.~Schieben, and N.~Merat, ``Defining interactions: A conceptual framework for understanding interactive behaviour in human and automated road traffic,'' {\em Theoretical Issues in Ergonomics Science}, vol.~21, pp.~728--752, Nov. 2020.

\bibitem{moralesAutomatedSynthesisNormative2013}
J.~Morales, M.~{Lopez-Sanchez}, J.~A. {Rodriguez-Aguilar}, M.~Wooldridge, and W.~Vasconcelos, ``Automated synthesis of normative systems,'' in {\em Proceedings of the 2013 International Conference on Autonomous Agents and Multi-Agent Systems}, {{AAMAS}} '13, (Richland, SC), pp.~483--490, {International Foundation for Autonomous Agents and Multiagent Systems}, 2013.

\bibitem{alonsoSpeedRoadAccidents2015}
F.~Alonso, C.~Esteban, C.~Calatayud, and {\`A}.~Egido, ``Speed and road accidents: Risk perception, knowledge and attitude towards penalties for speeding,'' {\em Psychofenia}, vol.~18, no.~31, pp.~63--76, 2015.

\bibitem{ohernKaahaajatFinnishAttitudes2023}
S.~O'Hern, V.~Vuorio, and A.~N. Stephens, ``Kaahaajat: {{Finnish Attitudes}} towards {{Speeding}},'' {\em International Journal of Environmental Research and Public Health}, vol.~20, p.~1995, Jan. 2023.

\bibitem{schafferOverdeterminingCauses2003}
J.~Schaffer, ``Overdetermining {{Causes}},'' {\em Philosophical Studies}, vol.~114, pp.~23--45, May 2003.

\bibitem{rudenkoHumanMotionTrajectory2020}
A.~Rudenko, L.~Palmieri, M.~Herman, K.~M. Kitani, D.~M. Gavrila, and K.~O. Arras, ``Human motion trajectory prediction: A survey,'' {\em The International Journal of Robotics Research}, vol.~39, pp.~895--935, July 2020.

\bibitem{kahnemanThinkingFastSlow2011}
D.~Kahneman, {\em Thinking, Fast and Slow}.
\newblock New York: {Farrar, Straus and Giroux}, 2011.

\bibitem{moralesSynthesisingLiberalNormative2015}
J.~Morales, M.~{L{\'o}pez-S{\'a}nchez}, J.~A. {Rodr{\'i}guez-Aguilar}, M.~Wooldridge, and W.~W. Vasconcelos, ``Synthesising liberal normative systems,'' in {\em Adaptive Agents and Multi-Agent Systems}, 2015.

\bibitem{pekVerifyingSafetyLane2017}
C.~Pek, P.~Zahn, and M.~Althoff, ``Verifying the safety of lane change maneuvers of self-driving vehicles based on formalized traffic rules,'' in {\em 2017 {{IEEE Intelligent Vehicles Symposium}} ({{IV}})}, (Los Angeles, CA, USA), pp.~1477--1483, IEEE, June 2017.

\bibitem{cavalcantesiebertMeaningfulHumanControl2022}
L.~Cavalcante~Siebert, M.~L. Lupetti, E.~Aizenberg, N.~Beckers, A.~Zgonnikov, H.~Veluwenkamp, D.~Abbink, E.~Giaccardi, G.-J. Houben, C.~M. Jonker, J.~{van den Hoven}, D.~Forster, and R.~L. Lagendijk, ``Meaningful human control: Actionable properties for {{AI}} system development,'' {\em AI and Ethics}, May 2022.

\bibitem{brahamAnatomyMoralResponsibility2012}
M.~Braham and M.~Van~Hees, ``An {{Anatomy}} of {{Moral Responsibility}},'' {\em Mind}, vol.~121, pp.~601--634, July 2012.

\bibitem{douerTheoreticalMeasuredSubjective2021}
N.~Douer and J.~Meyer, ``Theoretical, {{Measured}}, and {{Subjective Responsibility}} in {{Aided Decision Making}},'' {\em ACM Transactions on Interactive Intelligent Systems}, vol.~11, pp.~1--37, Mar. 2021.

\bibitem{mecacciResearchHandbookMeaningful2024}
G.~Mecacci, D.~Amoroso, L.~Cavalcante~Siebert, D.~Abbink, J.~van~den Hoven, F.~Santoni, and {Edward Elgar Publishing}, eds., {\em Research Handbook on Meaningful Human Control of Artificial Intelligence Systems}.
\newblock Cheltenham, UK ; Northampton, Massachusetts: Edward Elgar Publishing Limited, 2024.

\bibitem{santonidesioMeaningfulHumanControl2018}
F.~{Santoni de Sio} and J.~{van den Hoven}, ``Meaningful {{Human Control}} over {{Autonomous Systems}}: {{A Philosophical Account}},'' {\em Frontiers in Robotics and AI}, vol.~5, p.~15, Feb. 2018.

\bibitem{yazdanpanahDifferentFormsResponsibility2021}
V.~Yazdanpanah, E.~H. Gerding, S.~Stein, C.~Cirstea, M.~C. Schraefel, T.~J. Norman, and N.~R. Jennings, ``Different {{Forms}} of {{Responsibility}} in {{Multiagent Systems}}: {{Sociotechnical Characteristics}} and {{Requirements}},'' {\em IEEE Internet Computing}, vol.~25, pp.~15--22, Nov. 2021.

\bibitem{chenEvaluatingSafetyEfficiency2024}
K.~Chen, Z.~Li, P.~Liu, V.~L. Knoop, Y.~Han, and Y.~Jiao, ``Evaluating the safety and efficiency impacts of forced lane change with negative gaps based on empirical vehicle trajectories,'' {\em Accident Analysis \& Prevention}, vol.~203, p.~107622, Aug. 2024.

\bibitem{mullakkal-babuProbabilisticFieldApproach2020}
F.~A. {Mullakkal-Babu}, M.~Wang, X.~He, B.~Van~Arem, and R.~Happee, ``Probabilistic field approach for motorway driving risk assessment,'' {\em Transportation Research Part C: Emerging Technologies}, vol.~118, p.~102716, Sept. 2020.

\bibitem{jiaoUnifiedProbabilisticApproach2025}
Y.~Jiao, S.~C. Calvert, S.~Van~Cranenburgh, and H.~Van~Lint, ``A unified probabilistic approach to traffic conflict detection,'' {\em Analytic Methods in Accident Research}, vol.~45, p.~100369, Mar. 2025.

\end{thebibliography}

\end{document}